\documentclass{article}

\usepackage{arxiv}

\usepackage[utf8]{inputenc} 
\usepackage[T1]{fontenc}    
\usepackage{hyperref}       
\usepackage{url}            
\usepackage{booktabs}       
\usepackage{amsfonts}       
\usepackage{nicefrac}       
\usepackage{microtype}      
\usepackage{lipsum}
\usepackage{graphicx}
\graphicspath{{figures/}}

\usepackage[square,numbers]{natbib}
\usepackage{textcomp}
\usepackage{amsmath}
\usepackage{amssymb}
\usepackage{graphicx, subfigure}
\usepackage{array}
\usepackage[abs]{overpic}
\usepackage{tabularx} 
\usepackage{multirow}
\usepackage{hyperref}

\makeatletter

\usepackage{color}
\usepackage{soul}
\usepackage{amsthm}

\title{Elastostatics of star-polygon tile-based architectured planar lattices}

\author{
 Celal~Soyarslan\\
  Chair of Nonlinear Solid Mechanics \\ University of Twente, The Netherlands \\
  \texttt{c.soyarslan@utwente.nl} \\ 
   \And
 Andrew~Gleadall \\
  School of Mech., Elect. and Manuf. Eng.\\
  Loughborough University, UK\\
  \texttt{a.gleadall@lboro.ac.uk} \\
  \And
 Jiongyi~Yan \\
  School of Mech., Elect. and Manuf. Eng.\\
  Loughborough University, UK\\
  \texttt{j.yan@lboro.ac.uk} \\
  \And
 Hakan~Argeso\\
  Department of Aerospace Engineering\\
  Atilim University, Turkey \\
  \texttt{hakan.argeso@atilim.edu.tr} \\
  \And
 Emrah~Sozumert\\
  Zienkiewicz Centre for Comp. Eng.\\
  Swansea University, UK \\
  \texttt{emrah.sozumert@swansea.ac.uk}\\
}

\begin{document}
\maketitle

\begin{abstract}
A panoptic view of architectured planar lattices based on star-polygon tilings was developed. Four star-polygon-based lattice sub-families, formed of systematically arranged triangles, squares, or hexagons, were investigated numerically and experimentally. Finite-element-based homogenization allowed computation of Poisson's ratio, elastic modulus, shear modulus, and planar bulk modulus. A comprehensive understanding of the range of properties and micromechanical deformation mechanisms was developed. Adjusting the star-polygon angle achieved an over 250-fold range in elastic modulus, over a 10-fold range in density, and a range of $-0.919$ to $+0.988$ for Poisson's ratio. Additively manufactured lattices, achieved by novel printing strategies, showed good agreement in properties. Parametric additive manufacturing procedures for all lattices are available on  \url{www.fullcontrol.xyz/#/models/1d3528}. Three of the four sub-families exhibited in-plane elastic isotropy. One showed high stiffness with auxeticity at low density and a primarily axial deformation mode as opposed to bending deformation for the other three lattices. The range of achievable properties, demonstrated with property maps, proves the extension of the conventional material-property space. Lattice metamaterials with Triangle-Triangle, Kagome, Hexagonal, Square, Truncated Archimedean, Triangular, and Truncated Hexagonal topologies have been studied in the literature individually. Here, it is shown that these structures belong to the presented overarching lattice family.
\end{abstract}


\section{Introduction}
The exploitation of the structure-property link paves the path for the development of novel materials and structural design. To this end, there are two possible methods to apply. A structural modification is achievable by altering a material's chemical content with atomic or phase composition manipulations. An emerging trend is designing material geometry to create the so-called architectured materials \cite{ASHBY20134,shimada_hierarchical_2015}.  This, for instance, can be achieved by organising unit cell shapes in cellular materials, such as lattices, with various tessellations (hierarchical, stochastic, or periodic), element types (beam or surface), and connectivities (edge, face, or vertex) \cite{Bhateetal2019}.  Lattice materials are architectured materials built upon periodic arrangements of interconnected struts. By combining the merits of adjustable physical properties, low density, and functionality, they have been widely used in many applications spanning from  aerospace \cite{song_flutter_2017, cao_compression_2020, smeets_structural_2021,zheng_hierarchical_2013, mchale_morphing_2020} to medicine \cite{jiang_3d_2022, xiao_3d_2020}.  In recent works, desired macroscale material parameters (e.g., elastic and shear modulus) can be optimally satisfied by performing inverse parameter identification combined with genetic or topology optimization algorithms over lattice unit cell patterns \cite{Podesta2019, Yera2020, Yera2022, Rossi2021, Karathanasopoulos2022, Dosreis2022}.

Additive manufacturing expedites the trend of manufacturing architectured materials \cite{kim_designing_2019}, thanks to the ability to rapidly produce complex geometries with high precision without needing dedicated tooling or molds.  With digitally controlled tools for flexible structure design, additive manufacturing  has played an essential role in producing architectured lattices of inter-connected unit cells with intricate geometry \cite{SchaedlerCarter2016,Estrinetal2021, Dosreis2022}.  Traditional subtractive manufacturing of 2D planar lattices could introduce manufacturing defects into products, e.g., see, \cite{seiler_creep_2019, seiler_role_2019} for laser-cut metal honeycomb lattices and \cite{gu_experimental_2018} for water-jet cut triangular lattices. Additive manufacturing facilitates fabricating lattices with complicated geometry with varying material properties \cite{kim_designing_2019}. However, the conventional way of additive manufacturing by using computer-aided design (CAD) and slicing software to produce 2D lattices is challenging for intricate geometries. An inappropriate printing tool-path may result in inter-filament voids, unexpected geometry defects, and over- and under-extrusion, which can be detrimental to mechanical properties. A fully controlled printing tool-path and process are needed to achieve precise structures and to build reliable structure-property relationships.

In the design process of lattice structures, stiffness, strength, and Poisson's ratio are critical mechanical properties. In lattice materials, the effective mechanical properties depend on the slenderness ratio and the wall thickness of lattice struts \cite{zhang_fracturing_2018, dong_experimental_2019, Karathanasopoulos2018}. The stiffness and strength of hexachiral honeycomb-like lattice structures - namely, triangular, hexagonal, and Kagome lattices-with six-fold rotational symmetry were studied with numerical simulations and experiments on additively manufactured tough PLA \cite{tancogne-dejean_stiffness_2019} and  theoretical analysis on additively manufactured thermoplastic polymer ABS \cite{Cabras2014}.  Kagome structures (trihexagonal tilings) are popular in shape-morphing research due to their in-plane isotropy, high stiffness and strength, and low energy requirements for actuation applications \cite{pronk_quest_2017}. However, the controllability of deformation over the macroscopic lattice structure is a serious concern due to mechanical instabilities and structural defects \cite{wu_directing_2015}. The cell-wall thickness significantly affects the compression deformation mode of metallic auxetic re-entrant honeycomb lattices \cite{dong_experimental_2019}, and tensile strength is sensitive to imperfections in cell walls \cite{seiler_role_2019}. By controlling the density of cell walls, i.e., struts, mechanical and fracture properties of Maxwell lattices (twisted Kagome lattices) can be controlled \cite{zhang_fracturing_2018}. A hierarchical lattice is a combination of triangular, square, and hexagon lattices \cite{eddi_archimedean_2009}, and they host on-site symmetries, as well as sub-system symmetries (fine-grained symmetries) \cite{daniel_computational_2020}. Planar lattice systems such as square, triangular, and quasicrystal can be staggered to improve deformation performance in the direction of higher toughness, strength, and stiffness \cite{kim_designing_2019}.

Most engineering materials possess intrinsically positive Poisson's ratios (reduced size in the lateral direction perpendicular to the applied force-direction); e.g., Poisson's ratios for rubber, steel, gold, and various foams lie between 0.1 and 0.5. However, since the term auxetics was coined to describe materials with negative Poisson's ratios (increased thickness and width when being elongated in the longitudinal direction)\cite{EVANSetal1991}, studies have shown that certain natural or synthetic material microstructures at various scales may possess auxeticity  \cite{LAKES1038,Alderson1993,Nkansahetal1994,ChoiandLakes1995, Theocaris1997, MASTERS1996403,Grimaetal2001,Gaspar2011,BabaeeetalBertoldi2013}.  Indeed, energy arguments associated with the theory of elasticity allow negative Poisson's ratios. For 3D anisotropic elasticity, the Poisson's ratio is not bounded \cite{TingChen2005} whereas, for 3D isotropic elasticity, it varies from $-1$ to $1/2$.  2D rectangular crystals systems possess $\nu$ with  $\nu_\mathrm{max}\nu_\mathrm{min}<1$ whereas for 2D square and hexagonal crystal systems studied in this work, the bounds  $-1<\nu_\mathrm{min}\leq\nu_\mathrm{max}<1$ and $-1<\nu_\mathrm{min}=\nu_\mathrm{max}<1$ apply, respectively \cite{GAO2021104409}.  In contrast to planar materials possessing positive in-plane Poisson's ratios, which acquire a saddle shape upon out-of-plane bending, for planar auxetic materials, the emerging principal and transverse curvatures have identical signs to create synclastic curvature \cite{EvansAlderson2000,AldersonAlderson2007,Alderson1999}.  They find applications in textile, military, biomedical, and aerospace industries \cite{LI2018247,MetaImplants2018,MASTERS1996403}. Auxeticity also results in several superior mechanical properties, including indentation resistance, shear resistance \cite{EvansAlderson2000}, and plane strain fracture toughness \cite{ALDERSON19942261,EVANSetal1991}.  Poisson's ratios of modern metamaterials are extensively reviewed in \cite{Greavesetal2011NEW}. Most lattice materials with auxeticity are compliant, e.g., nonstandard microstructures constituting re-entrant load-bearing elements \cite{GibsonAshby1997,MASTERS1996403,ChoiandLakes1995,LAKES1038}. This feature makes lattice materials interesting and stimulates widespread research on them.

In this study, we investigate a class of architectured 2D lattice materials based on star-polygon tilings with an extensive range of structures and properties. By pursuing a combined numerical and experimental study, we demonstrate that for specific geometries, the analyzed planar lattices possess seemingly contrasting properties, such as relatively high elastic stiffness and auxeticity. Thanks to their architecture, they are able to extend the material property space \cite{SchaedlerCarter2016,ASHBY20134}. Detailed information about these lattices, e.g., unit cells as geometrical building blocks, associated symmetry classes and transformations, structural layouts for the selected internal lattice angles, chirality, elastomechanical planar symmetry, and auxeticity properties, is given in Figure\ \ref{fig:figure_major_1}. As demonstrated, all but M$_4$ acquire elastic isotropy. Generally, the lattices M$_2$, M$_3$, and M$_4$ are chiral, whereas M$_1$ is not. The lattice families M$_1$, M$_2$, M$_3$, and M$_4$ cover a wide range of geometries, a part of which are considered in the literature in an ad hoc sense. In Figure\ \ref{fig:figure_major_1}, these frequently studied lattices are highlighted with their  corresponding names. These are
Triangle-Triangle (TT), see, e.g., \cite{ hutchinson_structural_2006, shimada_hierarchical_2015, daniel_computational_2020, gu_experimental_2018, wu_directing_2015},
Kagome (K), see, e.g., \cite{kuang_bandgap_2005, hutchinson_structural_2006, zhang_mechanical_2008, eddi_archimedean_2009, elsayed_analysis_2010, wu_directing_2015, shimada_hierarchical_2015, niu_directional_2016, jeong_viscoelastic_2016, tankasala_tensile_2017, pronk_quest_2017, chen_study_2018, nelissen_2d_2019, zhu_theoretical_2019, daniel_computational_2020, liarte_multifunctional_2020, karathanasopoulos_latticemech_2020, tancogne-dejean_stiffness_2019,wu_directing_2015, zhang_fracturing_2018, fruchart_dualities_2020, liarte_multifunctional_2020},
Hexagonal (H), see, e.g., \cite{shimada_hierarchical_2015, daniel_computational_2020, pronk_quest_2017, karathanasopoulos_latticemech_2020, jeong_viscoelastic_2016, tancogne-dejean_stiffness_2019,eddi_archimedean_2009, mizzi_mechanical_2018, hou_mechanical_2018,seiler_creep_2019, vigliotti_mechanical_2013, vigliotti_nonlinear_2014, kim_designing_2019, seiler_role_2019, jiang_electromagnetic_2018, lei_nanoelectrode_2020, shabanpour_reconfigurable_2020, mousanezhad_hierarchical_2015, davami_ultralight_2015, cherkaev_damage_2019, xiang_yield_2018},
Square (S) see, e.g., \cite{daniel_computational_2020, tankasala_tensile_2017, elsayed_analysis_2010, zhang_mechanical_2008, jeong_viscoelastic_2016, eddi_archimedean_2009, hou_mechanical_2018, kim_designing_2019, jin_effect_2019, yuan_thermomechanically_2018}, Truncated Archimedean (AT), see, e.g., \cite{daniel_computational_2020, eddi_archimedean_2009},
Triangular (T) see, e.g., \cite{gu_experimental_2018, elsayed_analysis_2010, zhang_mechanical_2008, karathanasopoulos_latticemech_2020, tancogne-dejean_stiffness_2019, vigliotti_mechanical_2013, vigliotti_nonlinear_2014, kim_designing_2019, dos_reis_construction_2012, tankasala_crack_2020}, and, Truncated Hexagonal (TH), see, e.g., \cite{shimada_hierarchical_2015, daniel_computational_2020, eddi_archimedean_2009} lattices. This work allows us to treat these lattices in a unified framework and provide a comparison between different structures.
Star-shape perforations investigated in \cite{mizzi_mechanical_2018} result in rigid rotating triangle and square microstructures, constituting filled versions of some currently studied forms.

The paper has the following outline. Section\ \ref{S:theory} provides theoretical details regarding the geometrical properties of star-polygon tilings and the associated extended lattice family, periodic homogenization in planar lattices, and fundamental relations in 2D elasticity. Section\ \ref{S:resultsdiscussions} summarizes the numerical findings on finite and infinite lattice arrangements and their comparisons with experimental findings. This section also identifies the deformation modes of the lattices by studying bending and axial deformation energies. Finally, Section\ \ref{S:conclusion} draws conclusions.
\section{Theory}\label{S:theory}
\subsection{Geometric Properties of Star-Polygon Tilings and Associated Extended Family}
The family of lattices in the current study is based on uniform tilings by regular convex polygons and star-polygons, specifically star $3-$, $4-$, and $6-$gons \cite{GruenbaumShepherd1977}.
In Euclidean geometry, an equiangular and equilateral polygon is a regular $n-$gon whose internal angle amounts to $[n-2]\pi/n$. A star $n-$gon $\{n_\alpha\}$ is a 2D star-shaped (nonconvex) polygon with $n$ corners having
star-polygon angle $\alpha$ satisfying $0<\alpha<[n-2]\pi/n$, i.e., being smaller than the regular polygon interior angle. These corners are referred to as \emph{points} of the star, whereas the remaining corners, which are referred to as \emph{dents}, have angles $2[n-1]\pi/n-\alpha$.

\begin{figure*}[htbp]
   \centering
    \includegraphics[trim=15.mm 80.mm 15.mm 80.mm, clip,
  width=1.\textwidth]{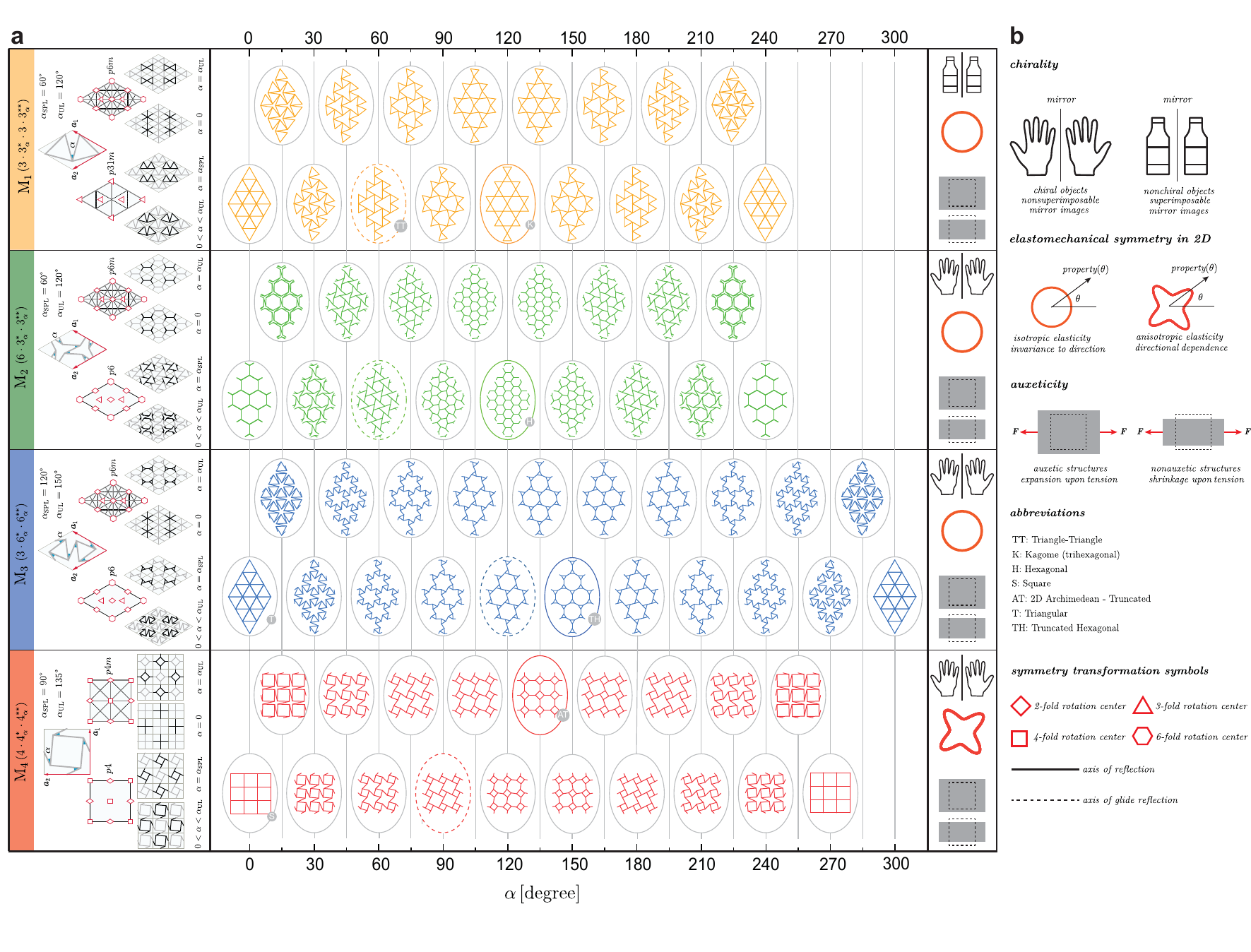}
  \vspace{-10mm}
  \caption{The geometric information, such as topology, selected periodic unit cell geometry, symmetry group, and the parametrization with star-polygon angle $\alpha$ thresholds for each lattice system M$_i$ for $i=1,\ldots,4$. Here, $\alpha_\mathrm{SPL}$ and $\alpha_\mathrm{UL}$ represent the star-polygon limit and uniqueness limit, respectively.  The symmetry elements used in diagrams detailing the crystallographic lattice group over unit cells are as follows: Bold lines show lines of reflection, whereas thin lines denote lines of a glide reflection. Unfilled rhombus symbols show the center of 2-fold rotation. Triangular, square, and hexagonal symbols indicate centers of 3-fold, 4-fold and 6-fold rotations. The crystallographic symmetry groups are given in terms of international symbols. Tilings for various $\alpha$ are shown. The lattices at $\alpha=\alpha_\mathrm{SPL}$ are distinguished with the dashed surrounding ellipse. The ones coincident with those studied in the literature in an ad hoc sense are marked with corresponding abbreviations. The properties of chirality, elastic anisotropy, and auxeticity are given on the right-hand column of a. In b, definitions of figures, symbols, and abbreviations are provided.
}
\label{fig:figure_major_1}
\end{figure*}

With this principle, all the four possible families of uniform tilings by regular convex polygons and star-polygons, whose topologies are identified as $4\cdot4_\alpha^*\cdot4_\alpha^{**}$, $3\cdot6_\alpha^*\cdot6_\alpha^{**}$, $6\cdot3_\alpha^*\cdot3_\alpha^{**}$ and $3\cdot3_\alpha^*\cdot3\cdot3_\alpha^{**}$, in which every corner is a vertex, are considered, with $\alpha<\alpha_\textrm{SPL}$ where $\alpha_\textrm{SPL}=[n-2]\pi/n$ denotes the star-polygon limit angle. For the sake of brevity, these topological structures shall be referred to as M$_1$, M$_2$, M$_3$, and M$_4$, respectively, see Figure\ \ref{fig:figure_major_1}. These topologies involve star $4-$, $6-$, $3-$, and $3-$gons with $\alpha_\textrm{SPL}$ being $90^\circ$, $120^\circ$, $60^\circ$ and $60^\circ$, respectively. M$_1$, M$_2$, and M$_3$ possess handedness, i.e., they are chiral, whereas M$_4$ does not possess this property.

This list is extended to include uniform tilings by regular convex polygons and non-star-polygons by considering $\alpha_\textrm{SPL}\leq\alpha<\alpha_\textrm{UL}$, where $\alpha_\textrm{UL}=[n-1]\pi/n$ is referred to as the limit $\alpha$ which provides geometrical uniqueness. Thus, $\alpha_\textrm{UL}-\beta$ and $\alpha_\textrm{UL}+\beta$ with $\beta<\alpha_\textrm{UL}$ creates geometrically identical structures for which only handedness is altered. For the four families, $\alpha_\textrm{UL}$ is computed as $135^\circ$, $150^\circ$, $120^\circ$ and $120^\circ$, respectively. The edges overlap for the lower limit with $\alpha=0$, and the lattice geometries degenerate to regular honeycombs composed of triangular, square, or hexagonal polygons. For the upper limit with $\alpha=\alpha_\textrm{UL}$, tilings by unimodal or bimodal regular convex polygons are generated. For both $\alpha=0$ and $\alpha=\alpha_\textrm{UL}$, the symmetry properties of the systems are enriched with mirror transformations where the handedness of the topologies M$_1$, M$_2$ and M$_3$ ceases to exist.

Each lattice sub-family is generated from the tiling of corresponding periodic unit cells (not unique), which is demonstrated in Figure\ \ref{fig:figure_major_1}. Tilings are realized together with translations applied to the periodic unit cell. The set of all such translations in 2D is written as $\{m \boldsymbol{a}_1+ n \boldsymbol{a}_2 \}$, where $\boldsymbol{a}_i$ for $i=1,2$ are the lattice basis vectors and $m$ and $n$ run independently through all integers, positive, negative and zero. Hence the unit cells are the building blocks of structures. In this sense, they encapsulate not only the geometrical properties but, from the point of view of micro-to-macro transformation with periodic homogenization, also the mechanical properties, which are intimately related.  M$_4$ lattice sub-family, which possesses elastic anisotropy, is a part of the square crystal system with chirality, whereas all the other lattice sub-families belong to the hexagonal crystal system with chirality and exhibit elastic isotropy.

\subsection{Periodic Homogenization in Planar Lattices} \label{S:homogenization}
The periodic homogenization framework presented here closely follows the previous work of the authors
\cite{soyarslan_argeso_bargmann_2018,SOYARSLAN2019280}. We consider continua with microstructure. Thus, the behavior of a typical material point at the macroscale is determined by a representative volume element (RVE) at the microscale. In the current periodic lattice systems, the RVE is equivalent to a periodic unit cell $\mathcal{V}$ whose volume\footnote{The volume $|\mathcal{V}|$ should not be confused with the volume of the solid phase within the unit cell.} and boundaries are denoted by $|\mathcal{V}|$ and $\partial\mathcal{V}$, respectively. Assuming constant strut thickness $\omega_{\mathrm{P}}$, the unit cell volume for the planar lattice is found by multiplying the unit cell area with the thickness.

We consider linear and infinitesimal planar elasticity. Let $\mathcal{B}$ denote the microscale referential configuration with $^\textrm{M}\mathcal{B}$ being its macroscopic counterpart.
The displacement field at $\boldsymbol{x}\in \mathcal{B}$ at time $t\in \mathcal{R}_+$ is represented by $\boldsymbol{u}(\boldsymbol{x},t)$. At the outset, the microscopic displacement gradient $\boldsymbol{H}$ can be computed with $\boldsymbol{H}=\boldsymbol{\nabla}\boldsymbol{u}$ where $\boldsymbol{\nabla}$ is the gradient operator. Letting $\boldsymbol e_i$ denote material base vectors and $\otimes$ the dyadic product operator,  $\boldsymbol{\nabla}\boldsymbol{u}=\partial u_i / \partial x_j \, \boldsymbol e_i \otimes \boldsymbol e_j$ for $i,j=1,2$. The symmetric part of $\boldsymbol{H}$ amounts to the microscopic strain tensor $\boldsymbol{\varepsilon}$ with $\boldsymbol{\varepsilon}:=\textrm{sym}(\boldsymbol{H})=1/2\,[\partial u_i / \partial x_j+\partial u_j / \partial x_i]\,\boldsymbol e_i\otimes \boldsymbol e_j$ for $i,j=1,2$. Thus, the strain tensor is symmetric with $\boldsymbol\varepsilon^\top=\boldsymbol\varepsilon$, with the superscript $\top$ implying the transpose.

Let $\boldsymbol{\sigma}$ denote the symmetric Cauchy stress tensor with  $\boldsymbol\sigma^\top=\boldsymbol\sigma$. Assuming plane stress state in $(\boldsymbol e_1,\boldsymbol e_2)$ plane, $\sigma_{33}=\sigma_{13}=\sigma_{23}=0$ and all the fields are independent of the out-of-plane coordinate $x_3$.  In the absence of dynamic effects and body forces, the corresponding microequilibrium equation reads $\textrm{div}\, \boldsymbol{\sigma}=\boldsymbol{0}$, where   $\textrm{div}$ is the divergence operator and $\boldsymbol 0$ the first-order zero tensor with $\textrm{div}\, \boldsymbol{\sigma}=\partial \sigma_{ij}/\partial x_j \, \boldsymbol e_i$ and $\boldsymbol{0}=0\, \boldsymbol e_i$, respectively, for $i,j=1,2$.

With the assumption of elastic isotropy at the microscale, the computation of the Cauchy stress $\boldsymbol{\sigma}$ is conducted via  Hooke's law
\begin{align}
\boldsymbol{\sigma}=\mathbb{C}:\boldsymbol{\varepsilon}\text{ in }\mathcal{B}\,.
\label{E:macrostress}
\end{align}
Here, $\mathbb{C}=\mathcal{C}_{ijkl}\, \boldsymbol e_i \otimes \boldsymbol e_j\otimes \boldsymbol e_k\otimes \boldsymbol e_l$ is the plane stress elastic constitutive tensor. With
$\Lambda$ and $\Upsilon$ denoting planar Lam\'{e} constants\footnote{Following relations link planar Young's modulus $E$, shear modulus $G$, bulk modulus (also referred to as area modulus) $K$ and Poisson's ratio $\upsilon$ to planar Lam\'{e} constants
\begin{equation}
E=
\dfrac{4\Upsilon[\Lambda+2\Upsilon]}{\Lambda+2\Upsilon},\,\text{ }G=\Upsilon,\,\text{ }
K=\Lambda+\Upsilon,\,\text{ }
\upsilon=\frac{\Lambda}{\Lambda+2\Upsilon}\,.
\end{equation}
Similarly, 3D elastic constants $E^{\mathrm{3D}}$, $G^{\mathrm{3D}}$, $K^{\mathrm{3D}}$ and $\nu^{\mathrm{3D}}$ can be derived from 3D Lam\'{e} constants $\Lambda^{\mathrm{3D}}$ and $\Upsilon^{\mathrm{3D}}$ which describes the 3D
elasticity tensor $\mathbb{C}^{\mathrm{3D}}$ whose components read $\mathcal{C}^{\mathrm{3D}}_{ijkl}=\Lambda^{\mathrm{3D}}\,\delta_{ij}\delta_{kl}+\Upsilon^{\mathrm{3D}}\,[\delta_{il}\delta_{jk}+\delta_{ik}\delta_{jl}]$ for $i,j,k,l=1,2,3$  with
\begin{equation}
E^{\mathrm{3D}}=
\dfrac{\Upsilon^{\mathrm{3D}}\,[3\,\Lambda^{\mathrm{3D}}+2\,\Upsilon^{\mathrm{3D}}]}{\Lambda^{\mathrm{3D}}+\Upsilon^{\mathrm{3D}}},\,\text{ }G^{\mathrm{3D}}=\Upsilon^{\mathrm{3D}},\,\text{ and }
K^{\mathrm{3D}}=3\,\Lambda^{\mathrm{3D}}+2\,\Upsilon^{\mathrm{3D}},\,\text{ and }
\upsilon^{\mathrm{3D}}=\frac{\Lambda^{\mathrm{3D}}}{2[\Lambda^{\mathrm{3D}}+\Upsilon^{\mathrm{3D}}]}\,.
\end{equation}
Using the relation $\Upsilon=\Upsilon^{\mathrm{3D}}\,,\, \Lambda=2\Lambda^{\mathrm{3D}}\Upsilon^{\mathrm{3D}}/[\Lambda^{\mathrm{3D}}+2\Upsilon^{\mathrm{3D}}]$, one can show that plane stress isotropic elasticity constants $E$, $G$ and $\nu$ are equal to their 3D counterparts with $E=E^\textrm{3D}$, $G=G^\textrm{3D}$ and $\nu=\nu^\textrm{3D}$, whereas $K\neq K^\textrm{3D}$ \cite{OstojaStarzewski2002}.} and $\delta_{ij}$ the Kronecker delta with $\delta_{ij}=1$ for $i=j$ and $0$ otherwise,
$\mathcal{C}_{ijkl}=\Lambda\,\delta_{ij}\delta_{kl}+\Upsilon\,[\delta_{il}\delta_{jk}+\delta_{ik}\delta_{jl}]$ for $i,j,k,l=1,2$.

At the macroscale, generalized Hooke's law is assumed with the assumption of a classical elastic continuum
\begin{align}
^\textrm{M}\boldsymbol{\sigma}=\mathbb{C}^\star:\, ^\textrm{M}\boldsymbol{\varepsilon}\text{ in } ^\textrm{M}\mathcal{B}\,.
\label{E:macrostress}
\end{align}
Here, the symmetry of stress and strain tensors at the microscale remains valid at the macroscale with $^\textrm{M}\boldsymbol\varepsilon^\top=$$^\textrm{M}\boldsymbol\varepsilon$ and  $^\textrm{M}\boldsymbol\sigma^\top=$$^\textrm{M}\boldsymbol\sigma$, where the macroscopic strain and stress tensors are respectively denoted by $^\textrm{M}\boldsymbol{\varepsilon}$ and $^\textrm{M}\boldsymbol{\sigma}$. $\mathbb{C}^\star$ denotes the planar effective elastic constitutive tensor.  In general, $\mathbb{C}^\star$ has nine effective constitutive constants, among which only six are linearly independent. Using $\gamma_{12}=2\varepsilon_{12}$, Eq.\ \eqref{E:macrostress} can be written in matrix form $[^\textrm{M}\boldsymbol{\sigma}]=[\mathbb{C}^\star][^\textrm{M}\boldsymbol{\varepsilon}]$ using Voigt notation as
\begin{align}
\left(
  \begin{array}{c}
    ^\textrm{M}\sigma_{11} \\
    ^\textrm{M}\sigma_{22} \\
    ^\textrm{M}\sigma_{12} \\
  \end{array}
\right)=
\left(
  \begin{array}{rrr}
    \mathcal{C}^\star_{1111} &  \mathcal{C}^\star_{1122} &  \mathcal{C}^\star_{1112} \\
    \mathcal{C}^\star_{1122} &  \mathcal{C}^\star_{2222} &  \mathcal{C}^\star_{2212}\\
    \mathcal{C}^\star_{1112} &  \mathcal{C}^\star_{2212} &  \mathcal{C}^\star_{1212}\\
  \end{array}
\right)
\left(
  \begin{array}{c}
    ^\textrm{M}\varepsilon_{11} \\
    ^\textrm{M}\varepsilon_{22} \\
    ^\textrm{M}\gamma_{12} \\
  \end{array}
\right)\,,
\label{E:macrostressVoigt}
\end{align}
with $\mathcal{C}^\star_{1111}$, $\mathcal{C}^\star_{1122}$, $\mathcal{C}^\star_{1112}$ $\mathcal{C}^\star_{2222}$, $\mathcal{C}^\star_{2212}$ and $\mathcal{C}^\star_{1212}$ denoting the six linearly independent elastic constants. Inverting Eq.\ \eqref{E:macrostress} gives  $^\textrm{M}\boldsymbol{\varepsilon}=\mathbb{S}^\star:\, ^\textrm{M}\boldsymbol{\sigma}$ in which $\mathbb{S}^\star=\mathcal{S}^\star_{ijkl}\, \boldsymbol e_i \otimes \boldsymbol e_j\otimes \boldsymbol e_k\otimes \boldsymbol e_l=\mathbb{C}^{\star\,-1}$ for $i,j,k,l=1,2$ denotes the elastic compliance tensor. In Voigt notation, this amounts to
\begin{align}
\left(
  \begin{array}{c}
    ^\textrm{M}\varepsilon_{11} \\
    ^\textrm{M}\varepsilon_{22} \\
    ^\textrm{M}\gamma_{12} \\
  \end{array}
\right)=
\left(
  \begin{array}{rrr}
    \mathcal{S}^\star_{1111} &  \mathcal{S}^\star_{1122} &  2\,\mathcal{S}^\star_{1112}\\
    \mathcal{S}^\star_{1122} &  \mathcal{S}^\star_{2222} &  2\,\mathcal{S}^\star_{2212}\\
    2\,\mathcal{S}^\star_{1112} &  2\,\mathcal{S}^\star_{2212} &  4\,\mathcal{S}^\star_{1212}\\
  \end{array}
\right)
\left(
  \begin{array}{c}
    ^\textrm{M}\sigma_{11} \\
    ^\textrm{M}\sigma_{22} \\
    ^\textrm{M}\sigma_{12} \\
  \end{array}
\right)\,.
\label{E:invertmacrostressVoigt}
\end{align}
The elasticity constants are determined by a computational homogenization process in which $i^\mathrm{th}$ (for $i=1,2,3$) column of the plane stress elastic stiffness matrix $[\mathbb{C}^\star]$
given in Eq.\ \eqref{E:macrostressVoigt} corresponds to the homogenized (macroscopic) stress tensor  $^\textrm{M}\boldsymbol{\sigma}^{\langle i \rangle}$ upon an imposed macroscopic strain tensor $^\textrm{M}\boldsymbol{\varepsilon}^{\langle i \rangle}$ with $^\textrm{M}\boldsymbol{\varepsilon}^{\langle 1 \rangle}=(1,0,0)^\top$, $^\textrm{M}\boldsymbol{\varepsilon}^{\langle 2 \rangle}=(0,1,0)^\top$ and $^\textrm{M}\boldsymbol{\varepsilon}^{\langle 3 \rangle}=(0,0,1)^\top$. The macroscopic strains $^\textrm{M}\boldsymbol{\varepsilon}^{\langle i \rangle}$ for $i=1,2,3$ are imposed on the RVE by employing a corresponding macroscopic displacement gradient $^\textrm{M}\boldsymbol{H}^{\langle i \rangle}$ through the selected control nodes. Considering the lattice arrangement and non-orthogonal RVE edges, two control nodes located at the corners of an imaginary unit square with positions
$\boldsymbol{x}^{\{1\}}=(1,0)^\top$ and $\boldsymbol{x}^{\{2\}}=(0,1)^\top$ are used to this end. The displacement vector of control node $j$ is fully prescribed with  $\boldsymbol{u}^{\{j\}\langle i \rangle}=$$^\textrm{M}\boldsymbol{H}^{\langle i \rangle}\cdot\boldsymbol{x}^{\{j\}}$. Letting  $\boldsymbol{x}^+\in\partial\mathcal{V}^+$ and $\boldsymbol{x}^-\in\partial\mathcal{V}^-$ denote two nodes periodically located at the periodic finite element model, the load cases are imposed under periodic boundary conditions considering periodic displacements $\boldsymbol{u}$ as\footnote{For planar beam finite element discretizations the periodicity of the rotational degree of freedom ${\Theta}$ is considered with $\Theta^{\langle i \rangle}(\boldsymbol{x}^+,t)-\Theta^{\langle i \rangle}(\boldsymbol{x}^-,t)=0$.}
\begin{equation}
\begin{split}
\boldsymbol{u}^{\langle i \rangle}(\boldsymbol{x}^+)-\boldsymbol{u}^{\langle i \rangle}(\boldsymbol{x}^-)&=\,^\textrm{M}\boldsymbol{H}^{\langle i \rangle}\cdot[\,\boldsymbol{x}^+-\boldsymbol{x}^-]\,.
\end{split}
\label{E:PBC}
\end{equation}
The uniqueness of the solution is guaranteed through constraining displacement at an internal node. With the limitation of geometrically linear analysis, $^\textrm{M}\boldsymbol{\sigma}_{\langle i \rangle}$  for $i=1,2,3$ are computed with \cite{Kouznetsova2001}
\begin{equation}
^\textrm{M}\boldsymbol{\sigma}^{\langle i \rangle}=\dfrac{1}{|\mathcal{V}|}\sum_{j=1}^2 \boldsymbol{f}^{\{j\}\langle i \rangle} \otimes \boldsymbol{x}^{\{j\}}\,,
\label{E:PBC2}
\end{equation}
where $\boldsymbol{f}^{\{j\}\langle i \rangle}$ is the reaction force at the control node $j$ during load case $i$. The symmetry of $^\textrm{M}\boldsymbol{\sigma}^{\langle i \rangle}$ is satisfied with the rotational equilibrium of forces $\boldsymbol{f}^{\{j\}\langle i \rangle}$ yielding $f_{2}^{\{1\}\langle i \rangle }=f_{1}^{\{2\}\langle i \rangle }$.
\subsection{Planar Elastic Moduli for $\mathrm{M}_1$, $\mathrm{M}_2$, $\mathrm{M}_3$, and $\mathrm{M}_4$}
Considering Eq.\ \eqref{E:macrostressVoigt}, possible symmetries involved in material geometry allow further reductions in the number of linearly independent material constants. Only four constants $\mathcal{C}^\star_{1111}$, $\mathcal{C}^\star_{1122}$, $\mathcal{C}^\star_{1112}$ and $\mathcal{C}^\star_{1212}$ are sufficient to prescribe planar elasticity of a material system with square symmetry. For this case $\mathcal{C}^\star_{1111}=\mathcal{C}^\star_{2222}$. With the nonzero shear-to-normal coupling terms with $\mathcal{C}^\star_{1112}=-\mathcal{C}^\star_{2212}\neq0$, a pure normal/shear strain creates shear/normal stresses. For hexagonal crystals, in-plane elastic isotropy is due, resulting in only two independent elasticity constants $\mathcal{C}^\star_{1111}$ and $\mathcal{C}^\star_{1212}$ where $\mathcal{C}^\star_{1122}=\mathcal{C}^\star_{1111}-2\mathcal{C}^\star_{1212}$. Here\footnote{For square and hexagonal crystals, the remaining linearly independent terms can be computed with only two - one normal and one shear - and only one -  superposition of normal and shear-load cases, respectively.}, the in-plane shear-to-normal coupling terms vanish with $\mathcal{C}^\star_{1112}=\mathcal{C}^\star_{2212}=0$. As a consequence, for the investigated lattices $\mathrm{M}_1$, $\mathrm{M}_2$, $\mathrm{M}_3$, and $\mathrm{M}_4$, the following definitions are used for the macroscopic planar Young's modulus $E_1^\star$ along direction $\boldsymbol e_1$, planar shear modulus $G_{12}^\star$, planar bulk modulus $K^\star$ and planar Poisson's ratio $\nu^\star_{21}$ for stretch direction $\boldsymbol e_1$ and probing direction $\boldsymbol e_2$, see, e.g., \cite{Jones1999,Lietal2019}
\begin{align}
E^\star_{1}=\dfrac{1}{\mathcal{S}^\star_{1111}}\,,\text{ }
G^\star_{12}=\dfrac{1}{4\,\mathcal{S}^\star_{1212}}\,,\text{ }
K^\star=\dfrac{1}{2\,[\mathcal{S}^\star_{1111}+\mathcal{S}^\star_{1122}]}\,,\text{ and }
\nu^\star_{12}=-\dfrac{\mathcal{S}^\star_{1122}}{\mathcal{S}^\star_{1111}}\,.
\label{E:all}
\end{align}
For in-plane isotropy, as in the case of hexagonal lattices, the elastic moduli $E^\star_{1}$, $G^\star_{12}$ and $\nu^\star_{12}$ defined in Eq.\ \ref{E:all} lose their directional dependence and corresponding subscripts are dropped.

In measuring the degree of planar elastic anisotropy of the lattices, the anisotropy index $A_{\textrm{SU}}$ given in \cite{Lietal2019} is used. $K^\star_{\textrm{V}}$ and $K^\star_{\textrm{R}}$ respectively denote Voigt and Reuss estimates of planar bulk moduli where $K^\star_{\textrm{V}}\geq K^\star_{\textrm{R}}$. $G^\star_{\textrm{V}}$ and $G^\star_{\textrm{R}}$ correspond to Voigt and Reuss estimates of shear moduli, respectively, with $G^\star_{\textrm{V}}\geq G^\star_{\textrm{R}}$. For lattices $\mathrm{M}_1$, $\mathrm{M}_2$, $\mathrm{M}_3$, and $\mathrm{M}_4$, the equivalence $K^\star_{\textrm{V}}=K^\star_{\textrm{R}}=K^\star$ (see Eq.\ \eqref{E:all}) holds which leads to
\begin{align}
A_{\textrm{SU}}=
\sqrt{2}\left[\dfrac{G^\star_{\textrm{V}}}{G^\star_{\textrm{R}}}-1\right]\,.
\label{E:anisotropy_index}
\end{align}
Here, $G^\star_{\textrm{V}}$ and $G^\star_{\textrm{R}}$ can be represented in terms of the elastic stiffness and compliance tensor components as
\begin{align}
G^\star_{\textrm{V}}=\dfrac{\mathcal{C}^\star_{1111}-\mathcal{C}^\star_{1122}+2\,\mathcal{C}^\star_{1212}}{4}\,\,\text{ and }\,\,
G^\star_{\textrm{R}}=\dfrac{1}{\mathcal{S}^\star_{1111}-\mathcal{S}^\star_{1122}+2\,\mathcal{S}^\star_{1212}}\,.
\label{E:anisotropy_indexv11}
\end{align}
For in-plane elastic isotropy
$A_{\textrm{SU}}=0$ with $G^\star_{\textrm{V}}=G^\star_{\textrm{R}}$. Otherwise, the larger the $A_{\textrm{SU}}$, the larger the  planar elastic anisotropy degree.
\section{Results and Discussions}\label{S:resultsdiscussions}
\subsection{Investigations on Periodic Unit Cells}
\begin{figure*}[htbp]
   \centering
    \includegraphics[trim=0.mm 70.mm 0.mm 70.mm, clip,
  width=1.\textwidth]{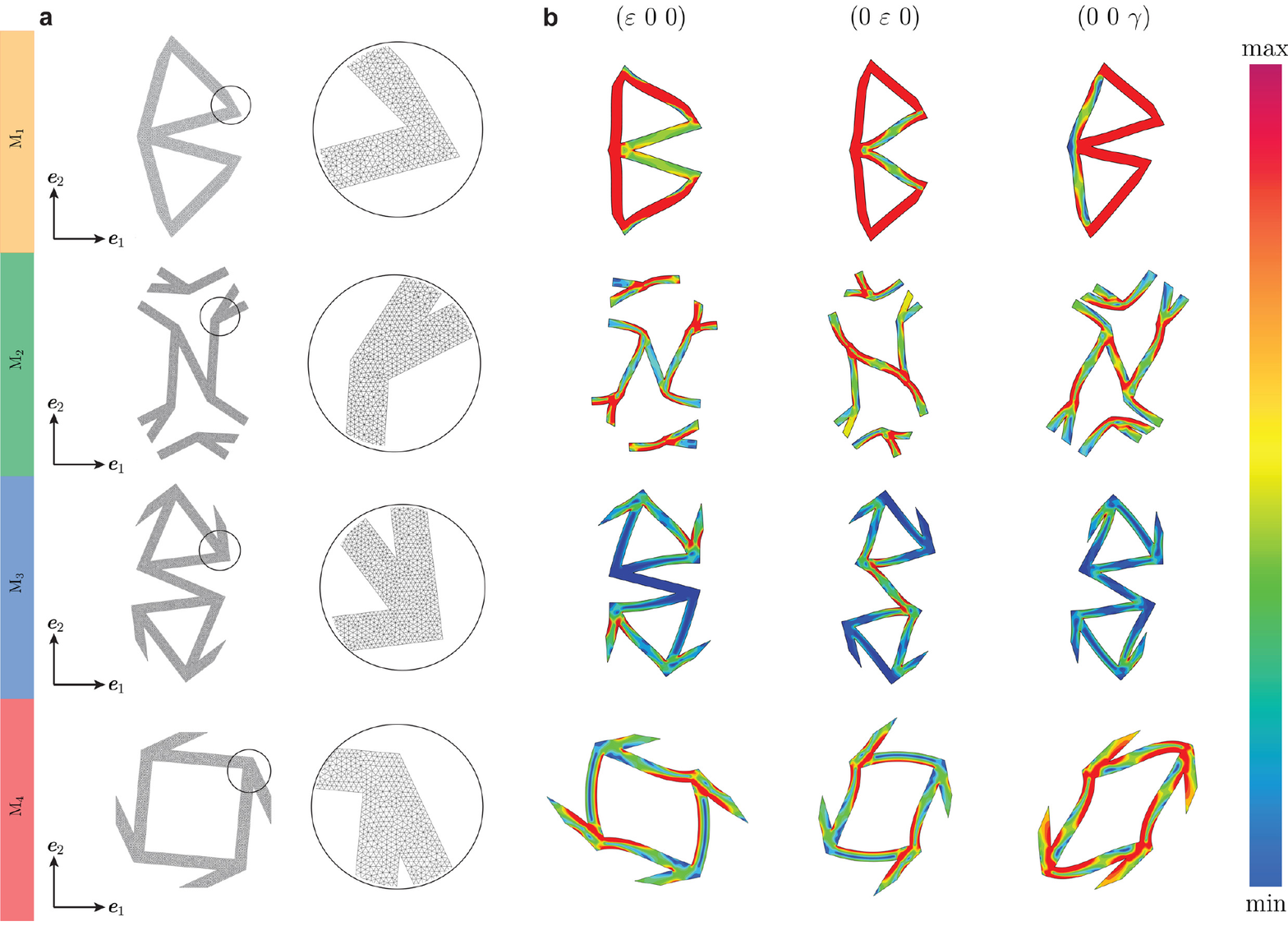}
  \vspace{-7mm}
  \caption{Finite element discretization (a) and von Mises stress plots for the considered strain-controlled loading cases identified by the given macroscopic strain component given above each column (two  plane strain normal and one shear) for M$_i$, $i=1,\ldots,4$, with $\alpha=30^\circ$ and $\lambda=30$. In the discretization, six-node quadratic plane stress continuum elements with an average size resulting in 75 elements along strut length are used. In (b), the macroscopic strain levels in each load case are identical for all lattice classes.
  For visualization purposes, a fixed stress threshold is assigned above the von Mises stress field and is plotted in red.
}
\label{fig:figure_mesh_and_stress}
\end{figure*}

Finite element analyses of frame structures with beam elements suffer from inaccuracy of connections at points to model strut-strut junctions with acute angles where $\alpha<30^\circ$.  Thus, unless otherwise stated, all the presented results are those of quasi-static implicit finite element simulations with \textsc{Abaqus} \cite{ABAQUS2019} six-node quadratic plane stress continuum elements with an average size resulting in 75 elements along the strut length, see Figure\ \ref{fig:figure_mesh_and_stress}. Accordingly, any out-of-plane deformation mode is suppressed. In effect, although the thickness changes are allowed, each node has two translational $(u,v)$ along $x-$ and $y-$directions, respectively.  The strut material is assumed to be isotropic and linear elastic. For the elastic material parameters for the solid phase $E_\mathrm{P}$ and $\nu_\mathrm{P}$, those corresponding to the raw polymer material used in the additive manufacturing process are selected ($E_\mathrm{P}=2700$ MPa and $\nu_\mathrm{P}=0.36$). Key numerical results of these investigations are tabulated in Tables\ from \ref{T:1} to \ref{T:9} in \ref{S:AppA}.
The lattices are described using three parameters. These are $\alpha$, strut length $L$ and thickness $\omega_{\mathrm{P}}$. The strut cross-sections are assumed to be square. Following Ref.\ \cite{Phani2006}, we link $L$ and $\omega_{\mathrm{P}}$ with the slenderness ratio defined as $\lambda=2\sqrt{3}L/\omega_{\mathrm{P}}$.

\begin{figure*}[htbp]
   \centering
  \includegraphics[trim=0.mm 70.mm 0.mm 70.mm, clip,
  width=1.\textwidth]{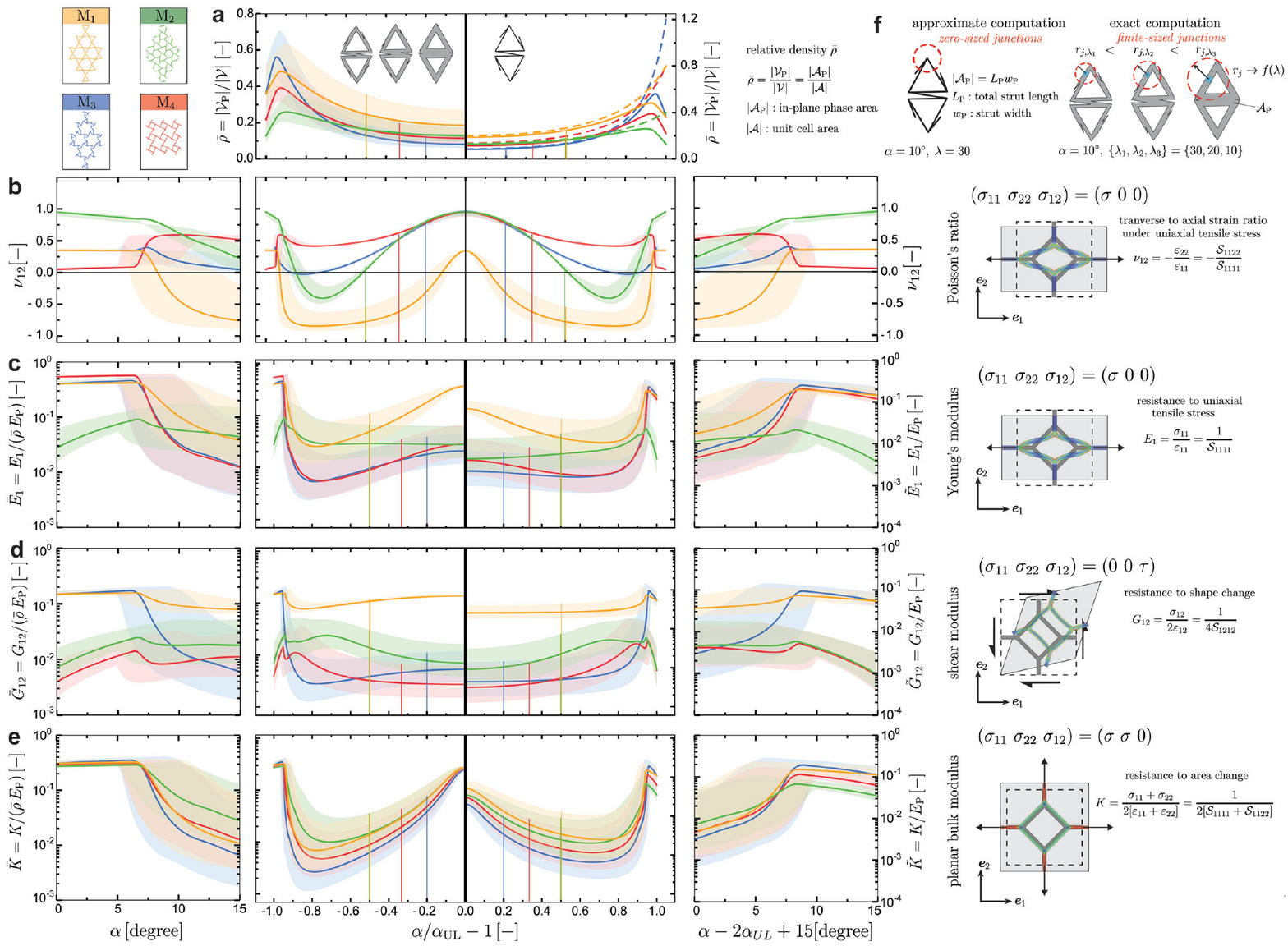}
  \vspace{-7mm}
  \caption{ The variation of the relative density (a) and effective elasticity parameters such as Poisson's ratio (b), Young's modulus (c), shear modulus (d), and planar bulk modulus (e), for the lattice systems M$_i$ for $i=1,\ldots,4$. The plots are symmetric for $\alpha_\mathrm {UL }$. The physical essence of each property is clarified using the exemplary loading modes associated with the computed parameter given on the right-hand side.  Three slenderness ratios are considered in the plots with $\lambda\in\{20,30,40\}$. The slenderness ratio change is provided by modifying the strut lengths by keeping the cross sections constant.  The semi-transparent bands cover $\lambda-$range, whereas the  lines  represent $\lambda=30$. Due to the periodic nature of the lattices, these results are equivalent to those for structures with unbounded size.  Young's, shear, and planar bulk moduli are scaled by dividing by relative density and Young's modulus of the base material, a polymer. For all plots except for the density, the intervals $0<\alpha<15^\circ$ and $2\alpha_\mathrm {UL }-15^\circ <\alpha<2\alpha_\mathrm {UL }$ are given in more detail on the left-hand side.}
  \label{fig:elasticmoduli_all}
\end{figure*}

Figure\ \ref{fig:elasticmoduli_all} summarizes the variation of the relative density\footnote{For the sake of conciseness, the $\star$ superscript used for highlighting a macroscopic quantity is dropped in the subsequent text and Figures.} as well as various elastic moduli as a function of $\alpha$ for the unit cells. For convenience in representations, $\alpha$ is scaled with $\alpha_\mathrm{UL}$. Due to the periodic nature of the lattices, these results are equivalent to those for structures with unbounded size. Three slenderness ratios $\lambda\in\{20,30,40\}$ are considered. The solid lines belong to $\lambda=30$, whereas the surrounding band forms the envelope for the given slenderness ratio interval. The slenderness ratio change is provided by modifying the strut lengths by keeping the cross sections constant. The sensitivity of the mechanical response to the strut slenderness reduces as $\alpha\to0$ and $\alpha\to\alpha_\mathrm{UL}$.

The results are demonstrated for the whole range $0 \leq \alpha \leq 2\alpha_\textrm{UL}$. All the curves show mirror symmetry with respect to $\alpha = \alpha_\textrm{UL}$. The vertical colored lines demonstrate the star-polygon angle limit $\alpha_\textrm{SPL}$ for the corresponding lattice of the same color. The relative density, see Figure\ \ref{fig:elasticmoduli_all}.a, and the elastic properties, see Figures\ \ref{fig:elasticmoduli_all}.b, c, d, and e, show rapid  fluctuations for the intervals $\alpha\in[0,15^\circ]$ and $\alpha\in[2\alpha_\mathrm{UL}-15^\circ,2\alpha_\mathrm{UL}]$. Thus, the variations of the mechanical properties at these regions are given in more detail in the additionally provided plots on the left and right of each row, respectively. A nondimensional representation is pursued by scaling Young's modulus, the shear modulus, and the bulk modulus with Young's modulus. The right-hand side plots are scaled only for the phase Young's modulus, whereas the ones on the left-hand side are also divided with the relative density. The former scaling factor aims at a distribution viable to other material classes with different elasticity moduli considering the negligible influence of the Poisson's ratio of the base material. The latter scaling aims to reflect material efficiency, a key factor to be considered in material selection. The higher the normalized magnitude, the higher the stiffness per unit volume of material.

In the relative density computations demonstrated in Figure\ \ref{fig:elasticmoduli_all}.a both 1D and 2D approaches are considered. The 1D computations have broken lines. Whereas the latter computation is geometrically exact, the former model beam approach is an approximation since the total in-plane phase area $A_\textrm{P}=L_\textrm{P}\,\omega_\textrm{P}$ is determined by summing up the in-plane strut lengths $L_\textrm{P}=\sum L$ within the depicted unit cells. The center-to-center connections in the beam assumption are not suitable for deep beams, i.e., beams with low slenderness ratios are considered. Since model beams have no thickness geometrically, the volumetric overlaps in the regions approaching the junctions are not considered. This error increases as $\alpha\to0$ and $\alpha\to2\alpha_\textrm{UL}$ for which the ligaments fold on each other and overlap, see Figure\ \ref{fig:elasticmoduli_all}.a. This causes unrealistic relative density computations beyond unity, see, e.g., Figure\ \ref{fig:elasticmoduli_all}.

It is generally observed that the computed nondimensional kinematic variables, e.g., Poisson's ratio, are less sensitive to model errors than the modulus of elasticity.
As demonstrated in Figure\ \ref{fig:elasticmoduli_all}.a, minimum relative density is observed at $\alpha=\alpha_\textrm{UL}$ for each lattice. All computations are symmetric to $\alpha_\textrm{UL}$.   Moreover, among all lattices, M$_1$ has the maximum relative density at $\alpha$, whereas the density of M$_3$ is at a minimum at this point.

\begin{figure*}[ht]
   \centering
   \includegraphics[trim=0.mm 70.mm 0.mm 70.mm, clip,width=1.\textwidth]{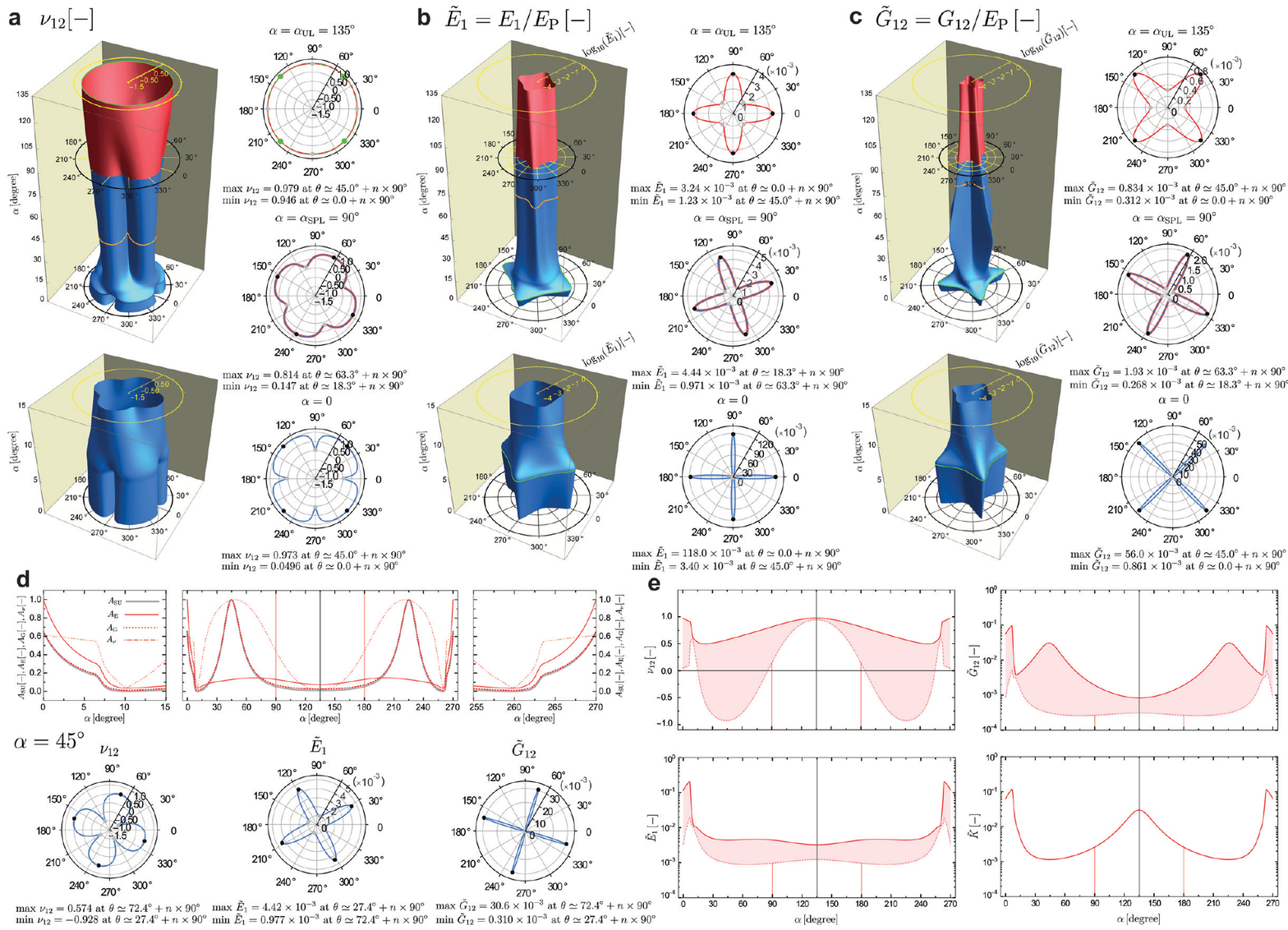}
  \vspace{-7mm}
  \caption{A detailed plot for the elastically anisotropic M$_4$. The directional distributions of Poisson's ratio (a) and Young's modulus (b), and shear modulus are plotted for $\alpha$ with $\lambda=30^\circ$ with a zoomed view for the interval $0<\alpha<15^\circ$. The sections of the 3D plots are explicitly given for the selected $\alpha$. (d) demonstrates the plot of various anisotropy indices as a function of $\alpha$. The index $A_\mathrm{SU}$ is given in Eq.\ \eqref{E:anisotropy_index}. For the remaining anisotropy indices, we have  $A_E=A_E^\alpha/\mathrm{max}(A_E^\alpha)$, $A_G=A_G^\alpha/\mathrm{max}(A_G^\alpha)$ and
  $A_\nu=A_\nu^\alpha/\mathrm{max}(A_\nu^\alpha)$ where $A_E^\alpha=E_\mathrm{max}/E_\mathrm{min}$, $A_G^\alpha=G_\mathrm{max}/G_\mathrm{min}$, and $A_\nu^\alpha=\nu_\mathrm{max}-\nu_\mathrm{min}$ for the given $\alpha$.  represent alternative anisotropy indices for the property denoted by the suffix at the section for the corresponding $\alpha$. In (e), the Poisson's ratio, Young's modulus, shear modulus, and planar bulk modulus plots are given with the maximum and the minimum values and the band representing intermediate values. As the planar bulk modulus is an invariant of the elasticity tensor, it is unique for each $\alpha$.}
  \label{fig:SP2_details}
\end{figure*}

The elastic moduli given in the remaining rows of Figure\ \ref{fig:elasticmoduli_all} are presented with reference to the Cartesian basis $(\boldsymbol e_1,\boldsymbol e_2)$. The relation of the unit cell base vectors to the Cartesian basis is given in Table\ \ref{T:basevectorcartesian}. In agreement with Neumann's principle, \cite{Neumann1885, Nye1985}, all the microstructures, except M$_4$, possess in-plane elastic isotropy. Therefore, for these structures, the results provided in Figure\ \ref{fig:elasticmoduli_all} do not depend on the directional choice; that is, they are valid for all in-plane directions. For M$_4$-type structures, however, this is not the case. Although the comparison to experiments is made possible with the choice of preferred orientation, many characteristics of this microstructure escape observation, e.g., auxeticity, are not observed. To bridge this gap, we provide the in-plane distribution of the Poisson's ratio, Young's, and shear moduli in Figure\ \ref{fig:SP2_details}. A detailed discussion regarding M$_4$-type structures is given in subsequent paragraphs.

\begin{table}[htbp]
\label{T:basevectorcartesian}
\caption{Direct lattice vectors $\boldsymbol{a}_i$ in terms of the Cartesian base vectors for the selected primitive unit cells, see Figure\ \ref{fig:figure_major_1}. The  edge length corresponding to each  unit cell $L_\textrm{UC}$ is computed with  $L_\textrm{UC}=|\boldsymbol{a}_1|=|\boldsymbol{a}_2|$.}
\centering
\begin{tabular}{ll}
\hline
lattice type & direct lattice vectors $\boldsymbol{a}_i$ for $i=1,2$\\
\hline
\multirow{2}{*}{M$_1$, M$_2$, M$_3$}
   & $\boldsymbol a_1=L_\textrm{UC}\cos{\pi/3}\, \boldsymbol e_1 + L_\textrm{UC}\sin{\pi/3}\, \boldsymbol e_2$\\
   & $\boldsymbol a_2=-L_\textrm{UC}\cos{\pi/3} \, \boldsymbol e_1 + L_\textrm{UC}\sin{\pi/3} \, \boldsymbol e_2$ \\
\hline
\multirow{2}{*}{M$_4$}
   & $\boldsymbol a_1=L_\textrm{UC}\,\boldsymbol e_1$ \\
   & $\boldsymbol a_2=L_\textrm{UC}\,\boldsymbol e_2$ \\
\hline
\end{tabular}
\end{table}

As demonstrated in the second row of Figure\ \ref{fig:elasticmoduli_all}, although the phase Poisson's ratio is 0.36, a wide variety of effective Poisson's ratios are computed for the microstructures, thanks to the structural layout of each lattice.
Still, lying within the interval $[-1,1]$, they satisfy the analytical bounds for 2D crystals.
For all lattice systems, a positive local maximum in the Poisson's ratio $\nu_{12}$ is observed at $\alpha=\alpha_\textrm{UL}$. Except for M$_1$, the observed local maxima are close to the upper theoretical bound. From $\alpha/\alpha_\textrm{UL}-1=0$ towards  $\alpha/\alpha_\textrm{UL}-1=\pm 0.6$, the Poisson's ratios show a monotonic decrease.
Down to the star-polygon angle limit, that is, for $\alpha_\textrm{SPL}<\alpha<\alpha_\textrm{UL}$, all structures remain nonauxetic except for M$_1$.
Thus, M$_1$ is the only structure that shows auxeticity outside the star range. Within the star range with $\alpha<\alpha_\textrm{SPL}$,  M$_2$ and M$_3$ also demonstrate auxetic behavior. For the selected probing direction, M$_4$ does not possess auxeticity. As tabulated in Table\ \ref{T:poisson_tabular}, among the investigated isotropic lattices, the smallest auxeticity occurs for M$_3$ with $\nu=-0.0287$ at $\alpha=28^\circ$. For the same lattice, maximum Poisson's ratio occurs at $\alpha\simeq\alpha_\textrm{UL}=150^\circ$ with a magnitude of $\nu=0.958$. This value corresponds to the maximum of the computed Poisson's ratios observed in the studied lattices. Unlike M$_3$, M$_1$ shows the largest auxeticity interval in $\alpha$. For M$_1$ maximum auxeticity occurs with $\nu=-0.844$ at $\alpha=30^\circ$. This corresponds to the minimum Poisson's ratio for the selected lattices. M$_1$, while being the only stretching-dominated lattice, shows the highest slenderness-ratio sensitivity in the Poisson's ratio for a wide range of $\alpha$. These results are due to
the alterations in the geometrical character of the junctions. For $\alpha=0$, M$_3$ and M$_1$ give the identical triangle lattice and thus identical Poisson's ratios. In Young's modulus distributions,  slenderness-ratio sensitivity is least at uniqueness angles and
symmetry points. These are the points where the structures behave mainly under stretching modes. In contrast, if bending is prominent, the slenderness-ratio sensitivity is greater. This observation is also compatible with general assumptions for Maxwellian lattices and their associated behavior.

In Figures\ \ref{fig:elasticmoduli_all}.c, d, and e, nondimensional Young's modulus, shear modulus, and planar bulk modulus distributions are demonstrated, respectively.
A local extremum in Young's moduli and bulk shear moduli is attained at $\alpha=\alpha_{\textrm{UL}}$ for all lattice systems except for M$_2$. For M$_2$, similar to its shear modulus, a local minimum of Young's modulus is observed at $\alpha=\alpha_{\textrm{UL}}$. Where all the other lattices show a wide variability of the scaled Young's modulus as a function of $\alpha$ for $30^\circ<\alpha<2\alpha_{\textrm{UL}}-30^\circ$, the plot for M$_2$ is nearly flat, signaling a small change with altering $\alpha$. This is counter to M$_2$'s Poisson's ratio response at the same interval. For the considered lattices, as $\alpha$ is gradually decreased from $\alpha=30^\circ$ there occurs a rapid increase in the scaled Young's moduli and planar bulk moduli. This trend is broken until a local maximum is obtained around $\alpha\simeq 8^\circ$.

Considering Young's modulus, shear modulus, and planar bulk modulus among the lattices, M$_1$ displays the stiffest elastic response for $\alpha$ out of the star range. At $\alpha=\alpha_{\textrm{UL}}$, at which all the lattices behave nonauxetic, Young's modulus and shear modulus of M$_1$ reach one and two orders of magnitude larger magnitudes than those of the other lattices. At $\alpha_{\textrm{UL}}$, the computed planar bulk moduli are relatively close. The lattices M$_3$ and M$_4$ behave mostly relatively compliant. Due to the anisotropy of M$_4$, this interpretation only corresponds to the selected probing direction. Auxetic materials are known to be highly compliant. M$_1$ constitutes the stiffest planar isotropic lattice even for the range where it possesses auxeticity when other lattices do not.

The 3D plots given in Figures\ \ref{fig:SP2_details}.a, b, and c demonstrate the directional dependence of Poisson's ratio, Young's modulus, and shear modulus as a function of $\alpha$ for M$_4$-type lattices respectively. Since the planar bulk modulus is invariant, it does not depend on the selected direction. Thus, its 3D plot is not given. The maximum and minimum values of Poisson's ratio, Young's modulus, shear modulus, and planar bulk modulus at each $\alpha$ are given in Figure\  \ref{fig:SP2_details}.e. Continuous and broken red lines represent maxima and minima, respectively. Intermediate values are marked with the colored region. Observed minimum Poisson's ratios demonstrated in Figure\  \ref{fig:SP2_details}.e reveals that M$_4$ behaves auxetically although not isotropically around the interval $15<\alpha<90$. The maximum observed Poisson's ratio occurs at $\alpha=135^\circ$ with $\nu=0.9794$ whereas the minimum Poisson's ratio is observed for $\alpha=44^\circ$ with $\nu=-0.9285$.

The polarity observed in the magnitudes seen in the demonstrated sections of the 3D plots in Figures\ \ref{fig:SP2_details}.a, b, and c shows the extent of elastic anisotropy in the material. For  $\alpha=\alpha_{\textrm{UL}}=135^\circ$ the anisotropy observed in Poisson's ratio diminishes with $\nu \in [0.9458,0.9794]$. These findings are supported by the anisotropy index $A_{\textrm{SU}}$ distribution given in Figure\ \ref{fig:SP2_details}.d at which the minimum magnitude is observed at $\alpha=\alpha_{\textrm{UL}}$. In contrast, the maximum anisotropy index $A_\mathrm{SU}$ is computed at $\alpha=45^\circ$. This corresponds to the point where the maximum of the stiffness component $c^\star_{13}$ is observed as a measure of chirality. The maximum chirality stiffness component provides the maximum anisotropy index.
The indices $A_E$, $A_G$, and $A_\nu$ represent the ratio of maximum and minimum of the observed property denoted by the suffix at the section for the selected $\alpha$. As demonstrated in Figures\ \ref{fig:SP2_details}.d, although $A_G$ denoting the anisotropy index relating to the shear modulus follows closely $A_\mathrm{SU}$, this is not the case for $A_E$, which represents the polarity of $E$.

Figure\ \ref{fig:figure_mesh_and_stress} demonstrates continuum finite element discretization and emerging stress fields in the corresponding lattice unit cells under the applied three strain-controlled loading cases and periodic boundary conditions. The von Mises stress plots reveal that the struts in lattice M$_1$ mainly experience an axial deformation mode, whereas those in the remaining lattices deform under bending. At junctions, stress concentrations develop. Considering that the developed stress field amounts to the resistance to applied macroscopic deformation, M$_3$ demonstrates relatively high compliance in both shear and normal loading behavior, which agrees with our previous observations. In contrast, M$_1$ is most resistant to deformation.

\subsection{Comparison with Experimental Findings and Simulations with Finite-Sized Samples}

\begin{figure*}[htbp]
   \centering
  \includegraphics[trim=0.mm 70.mm 0.mm 70.mm, clip,width=1.\textwidth]{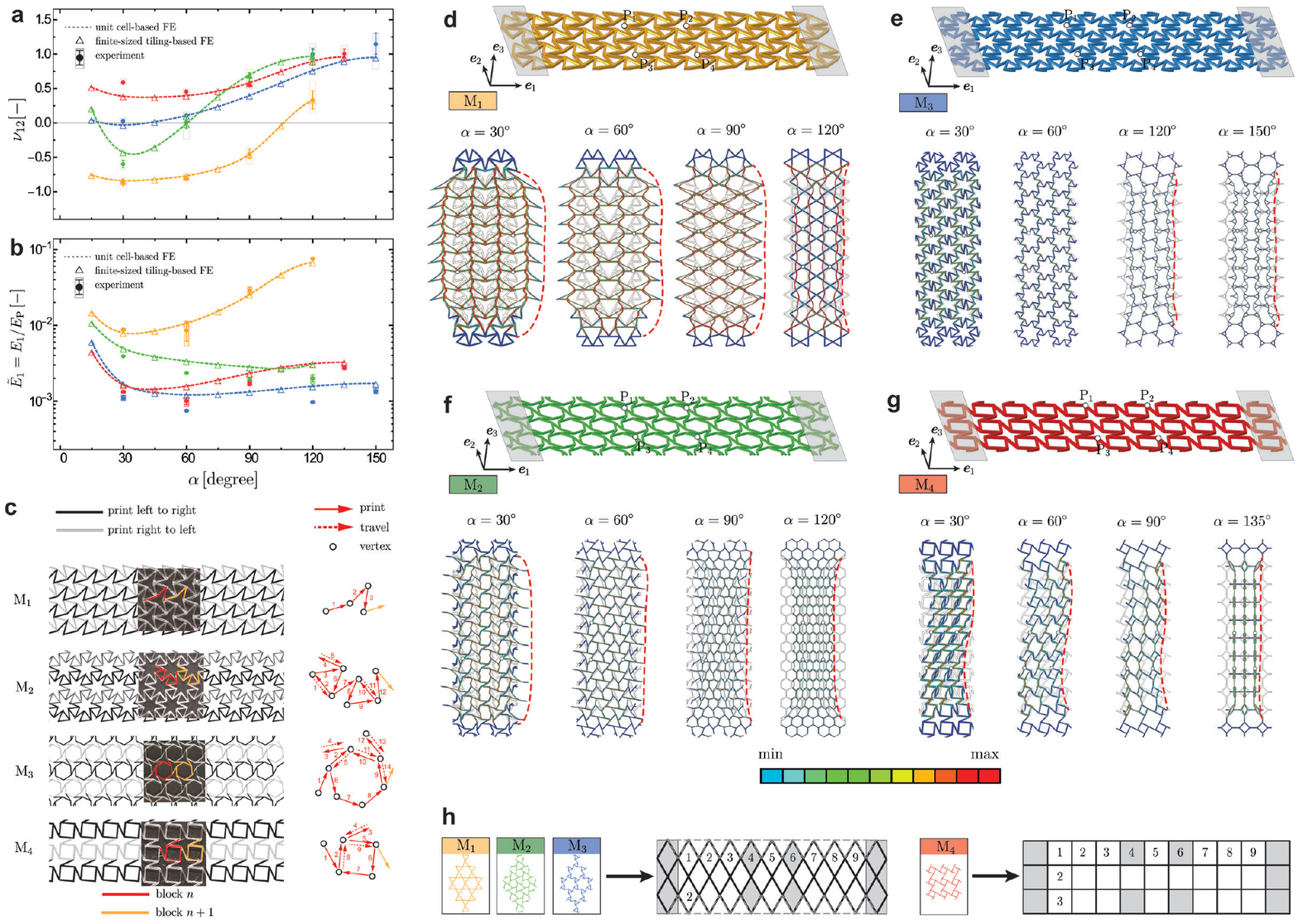}
  \vspace{-7mm}
  \caption{Results of experiments and simulations with finite-sized tilings. (a) and (b) display respectively the comparison of Poisson's ratios and Young's moduli. Here broken lines, hollow triangles, and filled markers with error bars and boxes represent unit cell-based FE results, finite-sized tiling-based FE results, and the experiments, respectively.  In the agreement between the periodic unit cell results and the finite-sized structural computations, selecting the finite-sized tiling is important. Here, the aspect ratio $\zeta$ of the unclamped region of $\zeta=1/\sqrt{3}$ was sufficient for M$_{i}$ with $i=1,2,3$ to have simulations with sufficiently suppressed size and boundary effects. For M$_{4}$, $\zeta=1/3$ was required.  Corresponding tile configurations, clamping, and sampling unit cells are depicted in (h). (d), (e), (f), and (g) demonstrate undeformed and deformed shapes with control nodes $\mathrm{P}_i$ for $i=1,\ldots,4$, which are sufficiently far away from the boundaries, used in kinematic computations exactly repeating the experimental practice, see Figure\ \ref{fig:A6}. In the demonstration of deformed shapes, axial displacements are suppressed, whereas transverse displacements are exaggerated for better visibility. The contour plot on the deformed mesh represents the equivalent strain field. In (c), a comparison with the simulation model and additively manufactured parts is given together with the information regarding printing paths.}
  \label{fig:3Dprint}
\end{figure*}

The M$_1$-M$_4$ lattices were additively manufactured for $\lambda=30$ and $\omega_\mathrm{P}=0.5$ mm at finite sizes with $\alpha=30^\circ$ to $150^\circ$. The printed lattice specimens were tensile tested, and Poisson's ratio was computed using the relative longitudinal and transverse deformations of selected control nodes on specimens, detailed in\ \ref{S:experimentalmethods}. These and the clamping region for each finite-sized lattice are shown in Figures\ \ref{fig:3Dprint}.d-h. Finite element analyses are conducted using the same finite-sized geometries and clamping conditions. Following the experiments, Young's modulus and Poisson's ratio were calculated using the measured tensile forces and control node kinematics.

Figures\ \ref{fig:3Dprint}.a and b demonstrate the comparison of unit cell-based FE model results (broken lines), finite-sized FE model results (hollow triangles), and the experiments (filled markers with error bars and boxes). For the experiments, the markers represent the mean value of three replicates, the error bars indicate the standard deviation, and the boxes indicate the extreme values (range). A good agreement can be found between the data points for experimental results and FE models. This provides confidence in the FE model's capabilities and the potential to translate theoretical findings to physical structures.

Each strut is printed as a single line from the nozzle; therefore, these structures represent the resolution limit achievable for this additive manufacturing process. To achieve good fidelity, it was necessary to explicitly define each movement of the nozzle. For this purpose, FullControl Gcode Designer \cite{GLEADALL2021102109} was used to create parametric designs where the unit cell size and quantity, $\alpha$ angle, and other parameters could be easily adjusted. Since the same parametric design was used for all structures within each M$_i$ type, any influences of the manufacturing procedure on experimental results were limited, which enabled consistent trends and good agreement with FE models. A common additive manufacturing workflow of creating a 3D CAD model and slicing it to retrospectively generate a tool-path would not achieve similar consistency. Therefore, any future work to implement these structures in practical applications should include the careful design of the tool-path. The manufacturing and tool-path design details are given in\ \ref{S:experimentalmethods}. Another key benefit of the parametric design approach is that it would be simple to implement transitions between different $\alpha$ angles at different positions within a structure and even to transition between the different structures M$_1-4$. The structures used in this paper can be parametrically adjusted and downloaded as manufacturing procedures from \url{www.fullcontrol.xyz/#/models/1d3528}.

The agreement between the periodic unit cell results and the finite-sized structural computations shows that the in-plane structural size according to Figure\ \ref{fig:3Dprint}.h, with an aspect ratio of $\zeta =1/3\sqrt{3}$ for M$_i$ for $i=1,2,3$ and $1/3$ for M$_4$,  is sufficient to avoid size and boundary effects. Figures\ \ref{fig:3Dprint}.a and b demonstrate that there is generally a good agreement between the experimentally observed and computed Poisson's ratios and
Figures\ \ref{fig:3Dprint}.d, e, f, and g depict the undeformed and deformed shapes for the finite-sized lattices in finite element simulations for selected $\alpha$. The contour plot shows the equivalent strain. For demonstration purposes, the displacements in the $x-$direction are exaggerated by a scale factor, whereas the $y-$direction displacements are omitted. Unlike other deformed shapes, those for the anisotropic lattice M$_4$ attain an S-like shape due to local rotations formed with chirality and the anisotropy whose major axes do not comply with the direction of the applied load for $\alpha=30^\circ$, $\alpha=60^\circ$ and $\alpha=90^\circ$. As the lattice is elastomechanically anisotropic, shear is invoked upon tension. To accommodate shear under applied constraints, the planar lattice deforms into an s-shape with local rotations thanks to the chiral microstructure. Once sufficient compliance is given by increasing the aspect ratio of the effective unclamped area of the structure, then the computed Poisson's ratios get closer to numerical results. This is inspiring in developing new lattice materials in arbitrary directions. For $\alpha=135^\circ$, the chirality is lost, and so is the S-shaped pattern with material rotations. Unlike M$_4$, a shear deformation mode is not invoked in isotropic lattices. Either a gradual bulging or contraction is observed at the gauge section for auxetic and nonauxetic lattices under the influence of supports constraining lateral contraction. Investigating the deformation contours, the auxetic behavior in the lattices M$_1$ and M$_2$ are observed for  $\alpha=30^\circ$, $\alpha=60^\circ$ and $\alpha=90^\circ$ for the former and $\alpha=30^\circ$ and $\alpha=60^\circ$ for the latter. For M$_3$, a slight auxeticity is observed only for $\alpha=30^\circ$. The findings agree with the auxeticity intervals given in Figure\ \ref{fig:elasticmoduli_all}.b.

\subsection{Deformation Modes: An Energetic Evaluation}
In energy computations, all lattices are considered rigid-jointed strut networks of beams which allow a clear separation of the deformation energy into  stretching, bending, and shear parts. With no prestress, each strut is discretized with 75 two-node 2D Timoshenko beam elements. Accordingly, each node has two translational $(u,v)$ along $x-$ and $y-$directions, respectively, and one rotational degree of freedom $\Theta_z$  about $z-$axis, which sum up to a total of three degrees of freedom. The total potential energy $\Phi$ per unit thickness along $z-$direction for a single strut is decomposed into stretching $\Psi_{\textrm{A}}$, bending $\Psi_{\textrm{B}}$ and shear $\Psi_{\textrm{S}}$ parts  with
\begin{equation}
\Psi=\Psi_{\textrm{A}}+\Psi_{\textrm{B}}+\Psi_{\textrm{S}}\,,
\end{equation}
where the components are given as
\begin{align}
\Psi_{\textrm{A}}&=\dfrac{1}{2}\int_{-L/2}^{L/2} E_\mathrm{P} \,\omega_\textrm{P}   \left[ \dfrac{\mathrm{d}u}{\mathrm{d}x} \right]^2 \mathrm{d}x\,,\\
\Psi_{\textrm{B}}&=\dfrac{1}{2}\int_{-L/2}^{L/2} E_\mathrm{P}\, I_z \left[ \dfrac{\mathrm{d}\Theta_z}{\mathrm{d}x} \right]^2 \mathrm{d}x\,,\\
\Psi_{\textrm{S}}&=\dfrac{1}{2}\int_{-L/2}^{L/2} \xi \,\omega_\textrm{P}\,  G   \left[ \dfrac{\mathrm{d}v}{\mathrm{d}x} - \Theta_z \right]^2 \mathrm{d}x\,.
\label{E:energies}
\end{align}
Here, $\omega_\textrm{P}$, $L$, and $I_z$, respectively, denote the width, length, and moment of inertia of the beam element. $\xi$ indicates the shear correction factor of the Timoshenko beam theory \cite{WeaverJohnston1987}.

\begin{figure*}[htbp]
   \centering
  \includegraphics[trim=0.mm 70.mm 0.mm 70.mm, clip, width=1.\textwidth]{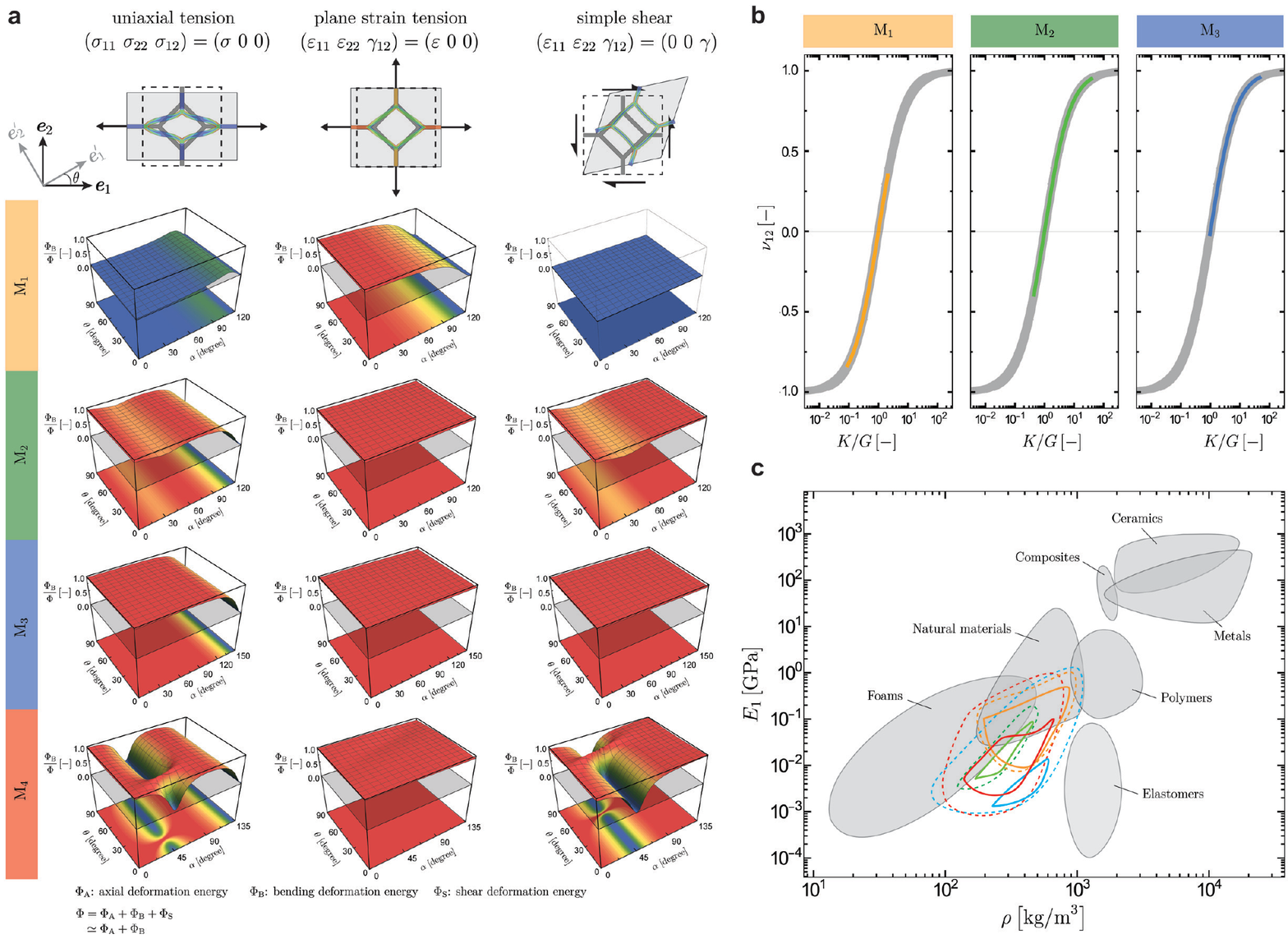}
  \vspace{-7mm}
  \caption{Bending to total deformation energy percentages for three macroscopically homogeneous loading scenarios: plane strain tension, uniaxial tension, and shear (a). $\lambda=60$ allowed a sufficiently suppressed shear deformation energy $\Phi_\mathrm{S}$. Poisson's ratio as a function of $K/G$ is given in (b). The color code identifies the lattice type as usual. The thick grey curve represents the plot for isotropic planar materials. In (c), the density-Young's modulus property maps for the newly introduced materials are added to the existing Ashby plots \cite{ASHBY20134} to demonstrate the extended property space with the introduction of the architectured planar lattices. Only computations within the range of selected slenderness ratios with $\lambda\in\{20,30,40\}$ are considered. The material properties used are those of the polymer. Here, the solid lines represent the boundary enclosing the results for only the auxetic lattices, whereas the dashed lines include all results.}
  \label{fig:energies}
\end{figure*}

Considering summations over all struts in a unit cell with
$\Phi_{\textrm{A}}=\sum\Psi_{\textrm{A}}$, $\Phi_{\textrm{B}}=\sum\Psi_{\textrm{B}}$, $\Phi_{\textrm{S}}=\sum\Psi_{\textrm{S}}$, and $\Phi=\sum\Psi$, one can compute the corresponding energy percentages for each deformation mechanism with
\begin{align}
\varphi_{\textrm{A}}=\dfrac{\Phi_{\textrm{A}}}{\Phi}\,, \quad
\varphi_{\textrm{B}}=\dfrac{\Phi_{\textrm{B}}}{\Phi}\,, \quad \text{and} \quad
\varphi_{\textrm{S}}=\dfrac{\Phi_{\textrm{S}}}{\Phi}.
\label{E:energypercentages}
\end{align}
For three macroscopically homogeneous loading scenarios, plane strain tension, uniaxial tension, and shear, the energy percentage plots are generated and demonstrated in Figure\ \ref{fig:energies}.a.
The reason for selecting these conditions is their use in the standard parameter identification practices for periodic unit cells.
In Figure\ \ref{fig:energies}.a, $\theta$ denote the considered loading direction with $\theta\in[0,90^\circ]$.
The analyses are realized for a slenderness ratio of $\lambda=60$, where shear deformation and junction effects are sufficiently suppressed. This makes two local deformation modes possible: stretching and shearing. These represent two ends of the rainbow plot as blue and red, respectively. Our simulations for $\lambda=30$ also show similar results. These plots demonstrate that the dominating deformation mode is bending for all lattices except for M$_1$. For the plane strain tension loading and as $\alpha$ tends to the uniqueness angle $\alpha_{\textrm{UA}}$, which is $150^\circ$ for M$_3$, $135^\circ$ for M$_4$ and $120^\circ$ for M$_2$, a rapid transition to an axial deformation mode is observed.
$\alpha_{\textrm{UA}}$ is the only angle the lattices M$_3$, M$_4$ and M$_2$ cease to be chiral. Thus, the results show that chirality mainly invokes bending energy modes for the lattices of the same topology.
Stored energy percentages over the elements are demonstrated on the exact figure for  $\alpha\to\alpha_{\textrm{UA}}$ and plane strain loading along $x-$direction.
The contours are given at the undeformed configuration.  It is seen that all members are mainly subject to a stretching mode, regardless of their location.

As all lattices except for M$_4$ are elastomechanically isotropic, they do not show any difference along the short edge of the plot. For M$_4$, however, which is an anisotropic lattice, as $\alpha\to45^\circ$ axial deformation mode is invoked for plane strain tension and shear loads. With the elastomechanical anisotropy of the lattice, around the loading direction of $\theta=20^\circ$, another deformation mode change is observed. Closer inspection reveals that for $\alpha=45^\circ$ this is the direction of the maximum macroscopic elastic compliance. It is known that for lattices, stretching deformation modes end up giving stiffer responses, which agrees with the drawn picture. The star-polygon angle threshold seems to have no observable influence on invoking different deformation energies.

Unlike other lattices, for the non-chiral M$_1$, a local stretching deformation mode prevails for plane strain tension and shear loading; see also Figure\ \ref{fig:figure_mesh_and_stress}. For uniaxial tension, bending domination is observed except for the case where $\alpha$ tends to the uniqueness angle $\alpha_{\textrm{UA}}=120^\circ$.

These results clearly show that four factors could affect the local deformation mode of lattice systems in addition to the lattice topology, otherwise known as member connectivity. These are strut slenderness ratio, member orientations or degree of chirality, and macroscopic loading type and direction. The former two influence the load distribution among members and how the applied loads are transferred among elements. The latter two are trivial; as for a beam element, an axial load will only invoke an axial deformation mode, whereas an applied couple invokes only a pure bending mode.

Figure\ \ref{fig:energies}.b demonstrates $K/G$ versus Poisson's ratio plot for the isotropic lattices only. The thick grey line in the background is the curve for elastically isotropic 2D materials for which $K/G=[1+\nu]/[1-\nu]$.
The fact that the curves for our isotropic lattices M$_1$, M$_2$, and M$_3$ lie on the grey line verifies our computations. The plot also allows investigation of the intervals for which the lattices are auxetic. M$_1$ is auxetic for most of the $K/G$ values, whereas M$_3$ shows auxeticity only for a small interval of $K/G$.

Material properties are mapped on the material chart relating density and Young's modulus given in Figure\ \ref{fig:energies}.c, \cite{ASHBY20134}. The mechanical properties of the star-polygon architectures extend the material property space for polymer materials by surpassing their conventional properties. Especially the M$_1$ lattice architecture enables lighter but stiffer structures which are also auxetic, demonstrated by the region with dashed enclosing boundaries. In contrast, most rigid and flexible polymer foams are not auxetic \cite{Rinde1970}.
\section{Conclusion}\label{S:conclusion}
This work studies the elastostatics of star-polygon tiling-based planar architectured lattice materials. The lattices are grouped into four sub-families M$_1$, M$_2$, M$_3$, and M$_4$. It is shown that many widely investigated lattice metamaterials that have been independently studied in the literature are specific variants of the presented overarching family. We conducted both numerical and experimental studies.

With an assumption of linear and infinitesimal Cauchy elasticity, numerical studies included first-order computational homogenization studies conducted over primitive periodic unit cells. This allowed computation of effective Poisson's ratio, Young's modulus, shear modulus, and planar bulk modulus for varying degrees of chirality and strut slenderness ratios.

Lattices are additively manufactured for selected parameters and subjected to tensile tests. The Poisson's ratios and elasticity moduli agree well with the simulations conducted on finite-sized structures and unit cells under periodic boundary conditions.

The lattices possess attractive properties in addition to their low relative densities. As its property maps prove,  the elastically isotropic M$_1$ extends
the material property space and demonstrates seemingly contrasting properties of relatively high stiffness in combination with auxeticity thanks to its axial primary deformation mode.
\section{Acknowledgements}
C.S. gratefully acknowledges Dr.~Swantje~Bargmann and the University of Wuppertal for the computational resources provided during the early developments of the work. E.S. gratefully acknowledges the funding in part from the EPSRC UK (grant number EP/R012091/1).
\clearpage
\appendix
\section{Experimental methods}\label{S:experimentalmethods}
Polylactic acid (PLA) (ECOMAX\textregistered, 3DXTECH US) was used to print all our lattice structures. Typical elongation at break is 5.2$\%$. The glass-transition temperature is 60\textdegree C. The recommended nozzle temperature is 210\textdegree C. Tensile tests of the lattice specimens were conducted to validate the finite-element simulations. PLA bulk-material properties were obtained from the material data sheet provided by 3DXTECH\textregistered, and used as input parameters for the simulations.

\subsection{Additive manufacturing of 2D lattice specimens}
Fullcontrol Gcode Designer \cite{GLEADALL2021102109}  was used for print path design and manufacturing Gcode generation, which facilitates parametric tool-path geometry design and process control. To minimize manufacturing defects, continuous printing paths were designed wherever possible for all lattices to avoid non-printing travel movements of the nozzle across the lattice since such movements cause defects. When an odd number of struts met at a junction (M$_2$, M$_3$, and M$_4$) it was not possible to simply print to and from the junction without double-printing one line. To avoid defects in such cases, novel intricate printing strategies were designed with material retraction, varied extrusion rate, and varied printing speed. Controlled non-printing movements were designed to occur along previously printed filaments without double-printing the struts, which would normally unintentionally increase their width. These carefully designed paths also avoided defects and disturbance to previously printed filaments since the nozzle did not cross any of them laterally but tracked within their edges. An odd number of unit cells were printed in the width direction to have complete configurations with periodic patterns that started and finished on the same side of the structure, meaning each layer could be repeated, as required for additive manufacturing, without excessive non-printing nozzle movement from the end of one layer to the beginning of the next. The nozzle temperature was 210\textdegree C. The filament spool was dried at 40\textdegree C before printing. A Raise3D Pro2\textregistered (Raise 3D Technologies, USA) printer was used. The printing speed was set to 1000 mm.min$^\textrm{-1}$, and the travelling speed was set to 6000 mm.min$^\textrm{-1}$. The extrusion width (printed filament width) was set to 0.5mm during GCode generation, the nozzle diameter was 0.4 mm, and the layer height was 0.167mm. All lattices were printed as 3-layer single-extrusion structures to achieve 0.5-mm thick struts.

\subsection{Tensile test and calculation of Poisson's ratio}
A tensile test was conducted on a universal tensile testing system (Instron\textregistered) with a load-cell of 2 kN and a constant strain rate of 0.03 min$^\textrm{-1}$ to obtain experimental material properties. For all specimens, three rows of the unit cells from both the bottom and top edges were clamped in the grips to avoid slipping or stress concentration. For Poisson's ratio, digital image correction (DIC) was used to measure longitudinal and transverse deformation. The tests were recorded by a high-resolution 4K camera to visualize and capture corrected local deformation, as seen in Figure A.6. Four unit cell corners (junctions) evenly spaced in 2 unit cells in the center of the specimens were marked with red dots (P$_1$-P$_4$), as the control nodes. This enabled the measurement of local strain by measuring the average change in L$_3$ and L$_4$ for longitudinal-direction deformation and the average change in L$_1$ and L$_2$ for transverse-direction deformation simultaneously. Twelve images were selected from the camera recordings for all specimens, corresponding to engineering strains starting from 0.0001 to yielding points within the linear elastic region, facilitating tracking of the longitudinal and transverse deformation with a constant engineering strain interval.

\begin{figure}[htb!]
   \centering
  \includegraphics[width= 0.48\textwidth]{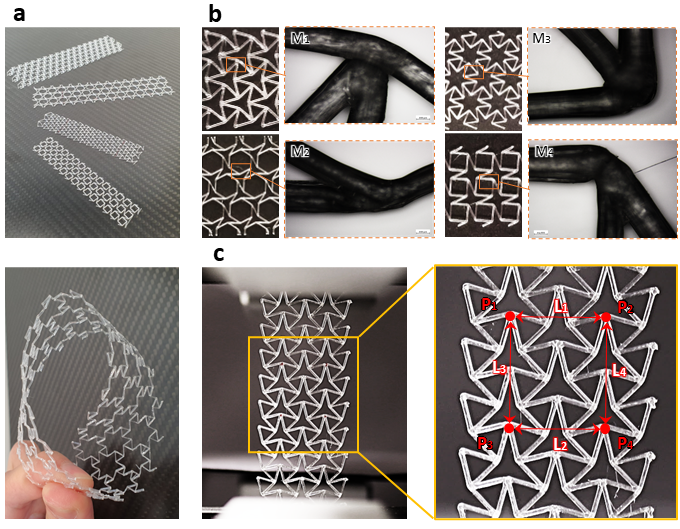}
  \caption{Additively manufactured planar lattices and tensile test (a). 3D printed lattices M$_1$ to M$_4$ with $\alpha=30^\circ$  and their micrographs in (b). (c)  Tensile test of M$_4$ ($\alpha=30^\circ$) specimens with a zoomed-in view for DIC for Poisson's ratio measurement.}
  \label{fig:A6}
\end{figure}

\section{The Influence of Junctions - A Comparison of Continuum and Beam Idealizations}

\begin{figure}[htb!]
   \centering
  \includegraphics[width= 0.48\textwidth]{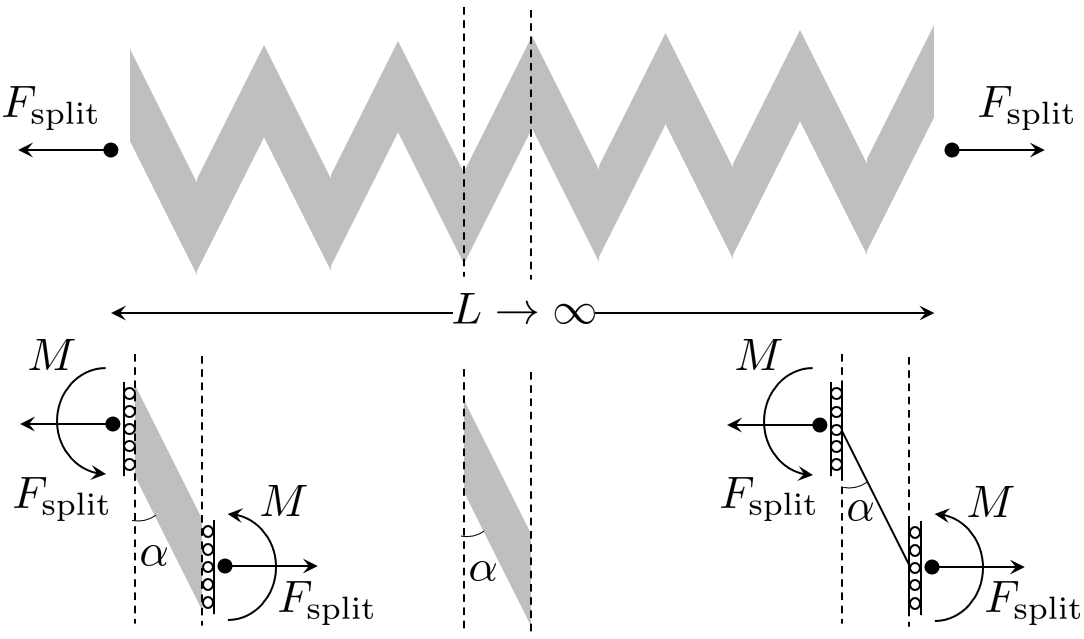}
  \caption{Designed splitting test for comparison of continuum and beam idealizations. In the continuum idealization, the domain is discretized into six-node quadratic plane stress continuum elements with an average size resulting in 75 elements along strut length. Beam idealization uses 75 planar Timoshenko (shear flexible) beam elements.}
  \label{fig:figure_split_test}
\end{figure}

In determining the junction effects and evaluating the validity range of beam element geometric parameters, a second test was designed, which is referred to as a splitting test.
This test extends an infinitely long, thin elastic spring with applied force at both ends. Considering the symmetry and translational periodicity of the problem, it is possible to concentrate on a repeating strut instead of the whole structure. Figure \ref{fig:figure_split_test} demonstrates the problem, its continuum, beam idealizations, and assigned boundary conditions.

\begin{figure}[htb!]
   \centering
  \includegraphics[width= 0.48\textwidth]{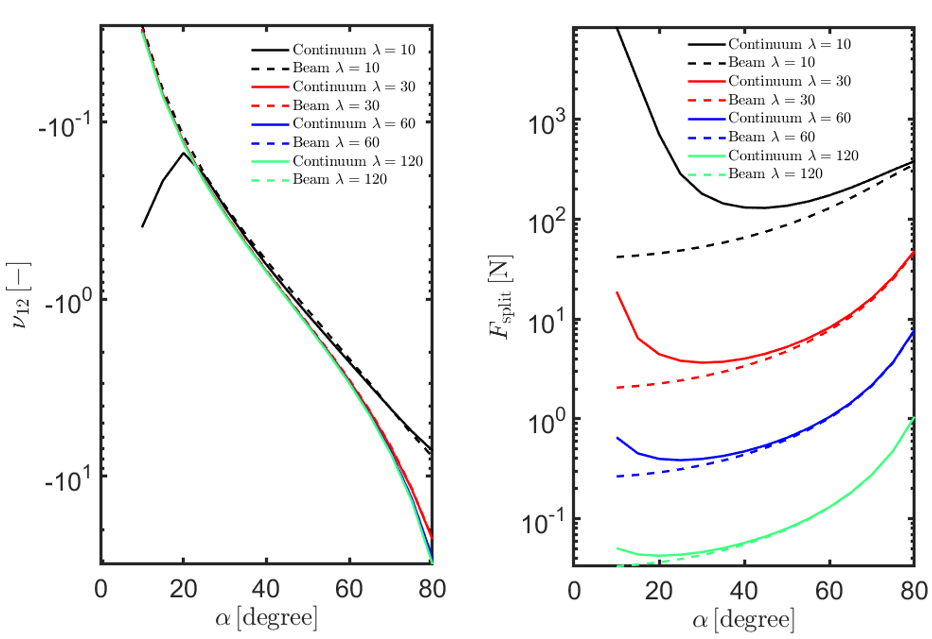}
  \caption{Result of the split test for $10\leq\lambda\leq120$. As seen, the difference in the computed Poisson's ratio  of the continuum and beam idealizations  is always smaller than that of the force. For the slenderness ratio is larger than 30 and $\alpha>30^\circ$, the results of the continuum and beam idealizations agree with each other.}
  \label{fig:figure_lambda_1}
\end{figure}

The numerical results are generated for two ranges of slenderness ratios. In Figure \ref{fig:figure_lambda_1}, Poisson's ratio and applied splitting force for  $10\leq\lambda\leq120$ is given. The difference in Poisson's ratios is generally smaller than in force measurements
 between the beam and continuum-based modeling approaches. Moreover, for the selected slenderness ratio, for $\alpha< 30^\circ$ the continuum and beam elements give increasingly different results.

\begin{figure}[htb!]
   \centering
  \includegraphics[width= 0.48\textwidth]{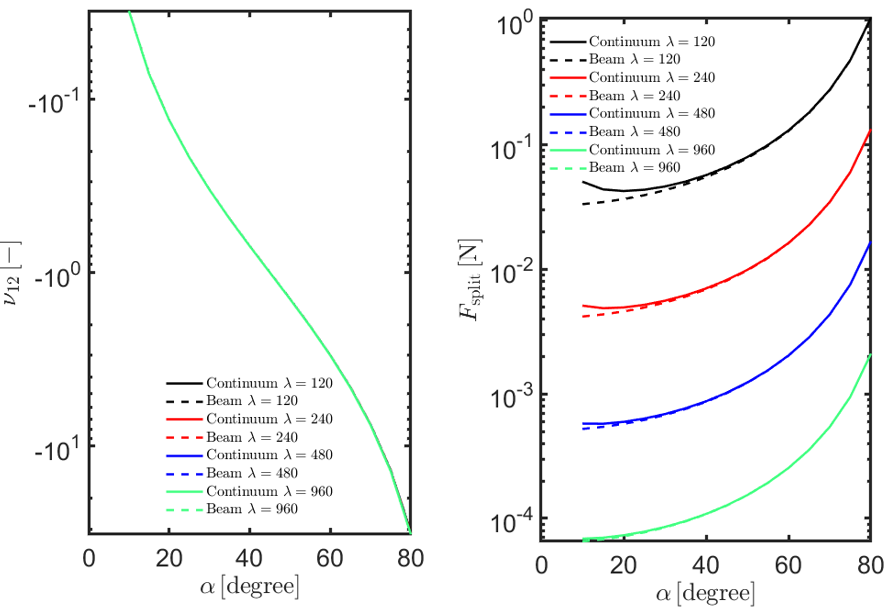}
  \caption{Result of the split test for $120\leq\lambda\leq960$. As seen, the difference in the computed Poisson's ratios of the continuum and beam idealizations  is always smaller than that of the force.}
  \label{fig:figure_lambda_2}
\end{figure}

In \ref{fig:figure_lambda_2}, with the increase of slenderness ratios, the difference in Poisson's ratios seem remedied between continuum and beam-based models, whereas for force measurements, there still is a difference for  $\alpha< 30^\circ$  and slenderness ratio of $120^\circ$

\section{Tabulated Data}\label{S:AppA}
The tables in this section present the values of relative density and material properties at the specified $\alpha $ values for the four families of uniform tilings having different slenderness ratios $\lambda $. Our computations are accomplished in two steps. In the first step, $\alpha $ is varied in between  $0^{\circ }$ and $15^{\circ }$ with the increments of $\Delta \alpha =0.1^{\circ }$, whereas in the second step, $\alpha $ is varied in between $15^{\circ }$ and $\alpha _\mathrm{UL}^{\circ }$ with the increments of $\Delta \alpha =1.0^{\circ }$. At each alpha value, $\theta $ is varied from $0^{\circ }$ to $360^{\circ }$ with increments of $\Delta \theta =0.1^{\circ }$. The values of any quantity, $\square $, at $\alpha =0$, $\alpha =\alpha _\mathrm{UL}$ and $\alpha =\alpha _\mathrm{SPL}$
are denoted by $\left[ \square \right] _{0}=\left. \square \right\vert_{\alpha =0^{\circ }}$, $\left[ \square \right] _\mathrm{UL}=\left. \square\right\vert _{\alpha =\alpha _\mathrm{UL}}$ and $\left[ \square \right]_\mathrm{SPL}=\left. \square \right\vert _{\alpha =\alpha _\mathrm{SPL}}$, respectively. For the chiral uniform tilings M$_{1}$, M$_{2}$ and M$_{3}$, which show isotropy, the maximum and minimum values of any material property, $\square$, with the corresponding $\alpha$ value these extrema appear, are denoted
by $\left[ \square \right] _{\max }\left\{ \alpha \right\} $ and $\left[\square \right] _{\min }\left\{ \alpha \right\} $. In case of M$_{4}$, these extrema are denoted by $\left[ \square \right] _{\max }\left\{ \alpha \right\} \left\langle \theta \right\rangle $ and $\left[ \square \right]_{\min }\left\{ \alpha \right\} \left\langle \theta \right\rangle $. Here $\left\langle \theta \right\rangle $ signifies the values of $\theta $ at which these extrema occur. Since M$_{4}$ does not possess isotropy, for a specified $\alpha $ value, Poisson's ratio, Young's modulus, and shear modulus are varied as a function of $\theta $. Therefore, in the relevant tables, material properties have maxima and minima with respect to $\theta$ for a given $\alpha$. On the other hand, in the case of bulk modulus values of M$_{4}$, this $\theta $ dependence vanishes. We also note
that, as discussed in the paper, two different types of normalizations are used in the presentation of the results of $E_{1}$, $G_{12}$ and $K$\, which are signified by putting "$-$" and "$\sim $" over the associated quantity.

\begin{table*}[htbp]
\caption{Relative density $\bar{\rho}$ values for the four families of uniform tilings having different slenderness ratios $\lambda $. The values of $\bar{\rho}$,
	 calculated by using continuum and beam elements in FEM analysis are denoted by $\bar{\rho}_{\mathrm{c}}$ and $\bar{\rho}_{\mathrm{b}}$, respectively. In the table, the values of $\bar{\rho}$ are multiplied by $10^{3}$  for representation purposes.}
\label{T:1}	
\centering
\resizebox{\textwidth}{!}{	
\begin{tabular}{llllllllllll}
	\hline
	& $\lambda $ & $\left[ \bar{\rho}_{\mathrm{c}}\right] _{0}$ & $\left[ \bar{\rho}_{\mathrm{b}}%
	\right] _{0}$ & $\left[ \bar{\rho}_{\mathrm{c}}\right] _\mathrm{UL}$ & $\left[ \bar{\rho}_{\mathrm{b}}%
	\right] _\mathrm{UL}$ & $\left[ \bar{\rho}_{\mathrm{c}}\right] _\mathrm{SPL}$ & $\left[ \bar{\rho}%
	_{b}\right] _\mathrm{SPL}$ & $\left[ \bar{\rho}_{\mathrm{c}}\right] _{\max }\left\{ \alpha
	\right\} $ & $\left[ \bar{\rho}_{\mathrm{b}}\right] _{\max }\left\{ \alpha \right\} $
	& $\left[ \bar{\rho}_{\mathrm{c}}\right] _{\min }\left\{ \alpha \right\} $ & $\left[
	\bar{\rho}_{\mathrm{b}}\right] _{\min }\left\{ \alpha \right\} $ \\
	\hline
	& 20 & 510 & 1200 & 270 & 300 & 355 & 400 & 633 $\left\{ 11.9^{\circ
	}\right\} $ & 1200 $\left\{ 0^{\circ }\right\} $ & 270 $\left\{ 120^{\circ
	}\right\} $ & 300 $\left\{ 120^{\circ }\right\} $ \\
	$M_{1}$ & 30 & 360 & 800 & 187 & 200 & 247 & 267 & 482 $\left\{ 9.1^{\circ
	}\right\} $ & 800 $\left\{ 0^{\circ }\right\} $ & 187 $\left\{ 120^{\circ
	}\right\} $ & 200 $\left\{ 120^{\circ }\right\} $ \\
	& 40 & 278 & 600 & 143 & 150 & 189 & 200 & 390 $\left\{ 7.7^{\circ }\right\}
	$ & 600 $\left\{ 0^{\circ }\right\} $ & 143 $\left\{ 120^{\circ }\right\} $
	& 150 $\left\{ 120^{\circ }\right\} $ \\
	&  &  &  &  &  &  &  &  &  &  &  \\
	& 20 & 190 & 600 & 190 & 200 & 240 & 257 & 353 $\left\{ 16^{\circ }\right\} $
	& 600 $\left\{ 0^{\circ }\right\} $ & 190 $\left\{ 0^{\circ },120^{\circ
	}\right\} $ & 200 $\left\{ 120^{\circ }\right\} $ \\
	$M_{2}$ & 30 & 129 & 400 & 129 & 133 & 164 & 171 & 260 $\left\{ 12.3^{\circ
	}\right\} $ & 400 $\left\{ 0^{\circ }\right\} $ & 129 $\left\{ 0^{\circ
	},120^{\circ }\right\} $ & 133 $\left\{ 120^{\circ }\right\} $ \\
	& 40 & 97.5 & 300 & 97.5 & 100 & 124 & 129 & 207 $\left\{ 10.4^{\circ
	}\right\} $ & 300 $\left\{ 0^{\circ }\right\} $ & 97.5 $\left\{ 0^{\circ
	},120^{\circ }\right\} $ & 100 $\left\{ 120^{\circ }\right\} $ \\
	&  &  &  &  &  &  &  &  &  &  &  \\
	& 20 & 510 & 1800 & 121 & 129 & 129 & 138 & 704 $\left\{ 10.3^{\circ
	}\right\} $ & 1800 $\left\{ 0^{\circ }\right\} $ & 121 $\left\{ 150^{\circ
	}\right\} $ & 129 $\left\{ 150^{\circ }\right\} $ \\
	$M_{3}$ & 30 & 360 & 1200 & 82.4 & 86.2 & 88.2 & 92.3 & 561 $\left\{
	7.9^{\circ }\right\} $ & 1200 $\left\{ 0^{\circ }\right\} $ & 82.4 $\left\{
	150^{\circ }\right\} $ & 86.2 $\left\{ 150^{\circ }\right\} $ \\
	& 40 & 278 & 900 & 62.5 & 64.6 & 66.9 & 69.2 & 467 $\left\{ 6.6^{\circ
	}\right\} $ & 900 $\left\{ 0^{\circ }\right\} $ & 62.5 $\left\{ 150^{\circ
	}\right\} $ & 64.6 $\left\{ 150^{\circ }\right\} $ \\
	&  &  &  &  &  &  &  &  &  &  &  \\
	& 20 & 316 & 1200 & 169 & 206 & 196 & 240 & 513 $\left\{ 12.6^{\circ
	}\right\} $ & 1200 $\left\{ 0^{\circ }\right\} $ & 169 $\left\{ 135^{\circ
	}\right\} $ & 206 $\left\{ 135^{\circ }\right\} $ \\
	$M_{4}$ & 30 & 218 & 800 & 115 & 137 & 133 & 160 & 391 $\left\{ 9.8^{\circ
	}\right\} $ & 800 $\left\{ 0^{\circ }\right\} $ & 115 $\left\{ 135^{\circ
	}\right\} $ & 137 $\left\{ 135^{\circ }\right\} $ \\
	& 40 & 166 & 600 & 86.8 & 103 & 101 & 120 & 317 $\left\{ 8.2^{\circ
	}\right\} $ & 600 $\left\{ 0^{\circ }\right\} $ & 86.8 $\left\{ 135^{\circ
	}\right\} $ & 103 $\left\{ 135^{\circ }\right\} $\\
	\hline
\end{tabular}

}

\end{table*}

\begin{table*}[htbp]
	\caption{Poisson's ratio $\nu _{12}$ values for the chiral uniform tilings having different slenderness ratios $\lambda $.}
	\label{T:poisson_tabular}
\centering		
\begin{tabular}{lllllll}
	\hline
	& $\lambda $ & $\left[ \nu _{12}\right] _{0}$ & $\left[ \nu _{12}\right]
	_\mathrm{UL}$ & $\left[ \nu _{12}\right] _\mathrm{SPL}$ & $\left[ \nu _{12}\right] _{\max
	}\left\{ \alpha \right\} $ & $\left[ \nu _{12}\right] _{\min }\left\{ \alpha
	\right\} $ \\
	\hline
	& 20 & 0.346 & 0.338 & -0.547 & 0.346 $\left\{ 0.0^{\circ }\right\} $ &
	-0.630 $\left\{ 36^{\circ }\right\} $ \\
	$M_{1}$ & 30 & 0.350 & 0.337 & -0.774 & 0.350 $\left\{ 0.4^{\circ }\right\} $
	& -0.844 $\left\{ 30^{\circ }\right\} $ \\
	& 40 & 0.349 & 0.337 & -0.870 & 0.349 $\left\{ 2.2^{\circ }\right\} $ &
	-0.919 $\left\{ 28^{\circ }\right\} $ \\
	&  &  &  &  &  &  \\
	& 20 & 0.892 & 0.892 & 0.0898 & 0.892 $\left\{ 0^{\circ },120^{\circ
	}\right\} $ & -0.221 $\left\{ 37^{\circ }\right\} $ \\
	$M_{2}$ & 30 & 0.949 & 0.949 & 0.0405 & 0.949 $\left\{ 0^{\circ },120^{\circ
	}\right\} $ & -0.406 $\left\{ 34^{\circ }\right\} $ \\
	& 40 & 0.971 & 0.971 & 0.0176 & 0.971 $\left\{ 0^{\circ },120^{\circ
	}\right\} $ & -0.490 $\left\{ 33^{\circ }\right\} $ \\
	&  &  &  &  &  &  \\
	& 20 & 0.346 & 0.912 & 0.742 & 0.912 $\left\{ 150^{\circ }\right\} $ &
	0.0309 $\left\{ 32^{\circ }\right\} $ \\
	$M_{3}$ & 30 & 0.350 & 0.958 & 0.768 & 0.958 $\left\{ 150^{\circ }\right\} $
	& -0.0287 $\left\{ 28^{\circ }\right\} $ \\
	& 40 & 0.348 & 0.976 & 0.779 & 0.976 $\left\{ 150^{\circ }\right\} $ &
	-0.0594 $\left\{ 25^{\circ }\right\} $\\
	\hline
\end{tabular}%

\end{table*}

\begin{table*}[htbp]
	\caption{Young's modulus $E_{1}$ values for the chiral uniform tilings having different slenderness ratios $\lambda $. In the table, the values of $E_{1}$
		are multiplied by $10^{3}$  for representation purposes.}
	\label{T:3}		
\centering
\resizebox{\textwidth}{!}{	
\begin{tabular}{llllllllllll}
	\hline
	& $\lambda $ & $\left[ \bar{E}_{1}\right] _{0}$ & $\left[ \tilde{E}_{1}%
	\right] _{0}$ & $\left[ \bar{E}_{1}\right] _\mathrm{UL}$ & $\left[ \tilde{E}_{1}%
	\right] _\mathrm{UL}$ & $\left[ \bar{E}_{1}\right] _\mathrm{SPL}$ & $\left[ \tilde{E}_{1}%
	\right] _\mathrm{SPL}$ & $\left[ \bar{E}_{1}\right] _{\max }\left\{ \alpha \right\}
	$ & $\left[ \tilde{E}_{1}\right] _{\max }\left\{ \alpha \right\} $ & $\left[
	\bar{E}_{1}\right] _{\min }\left\{ \alpha \right\} $ & $\left[ \tilde{E}_{1}%
	\right] _{\min }\left\{ \alpha \right\} $ \\
	\hline
	& 20 & 450 & 230 & 385 & 104 & 108 & 38.3 & 495 $\left\{ 9.6^{\circ
	}\right\} $ & 309 $\left\{ 9.7^{\circ }\right\} $ & 75.0 $\left\{ 33^{\circ
	}\right\} $ & 34.6 $\left\{ 43^{\circ }\right\} $ \\
	$M_{1}$ & 30 & 401 & 144 & 366 & 68.3 & 49.0 & 12.1 & 423 $\left\{
	6.4^{\circ }\right\} $ & 198 $\left\{ 6.5^{\circ }\right\} $ & 25.3 $\left\{
	27^{\circ }\right\} $ & 9.15 $\left\{ 34^{\circ }\right\} $ \\
	& 40 & 380 & 105 & 357 & 50.9 & 27.1 & 5.12 & 392 $\left\{ 4.8^{\circ
	}\right\} $ & 146 $\left\{ 4.9^{\circ }\right\} $ & 11.6 $\left\{ 23^{\circ
	}\right\} $ & 3.45 $\left\{ 30^{\circ }\right\} $ \\
	&  &  &  &  &  &  &  &  &  &  &  \\
	& 20 & 61.0 & 11.6 & 61.0 & 11.6 & 65.2 & 15.6 & 177 $\left\{ 10.2^{\circ
	}\right\} $ & 59.1 $\left\{ 10.5^{\circ }\right\} $ & 61.0 $\left\{ 0^{\circ
	},120^{\circ }\right\} $ & 11.6 $\left\{ 0^{\circ },120^{\circ }\right\} $
	\\
	$M_{2}$ & 30 & 27.6 & 3.55 & 27.6 & 3.55 & 28.7 & 4.70 & 91.2 $\left\{
	6.6^{\circ }\right\} $ & 21.4 $\left\{ 6.7^{\circ }\right\} $ & 27.6 $%
	\left\{ 0^{\circ },120^{\circ }\right\} $ & 3.55 $\left\{ 0^{\circ
	},120^{\circ }\right\} $ \\
	& 40 & 15.5 & 1.51 & 15.5 & 1.51 & 15.8 & 1.97 & 54.5 $\left\{ 4.9^{\circ
	}\right\} $ & 9.87 $\left\{ 5.0^{\circ }\right\} $ & 14.0 $\left\{ 28^{\circ
	}\right\} $ & 1.51 $\left\{ 0^{\circ },120^{\circ }\right\} $ \\
	&  &  &  &  &  &  &  &  &  &  &  \\
	& 20 & 450 & 230 & 44.3 & 5.36 & 38.7 & 5.00 & 547 $\left\{ 8.9^{\circ
	}\right\} $ & 381 $\left\{ 8.9^{\circ }\right\} $ & 17.2 $\left\{ 43^{\circ
	}\right\} $ & 4.32 $\left\{ 75^{\circ }\right\} $ \\
	$M_{3}$ & 30 & 401 & 144 & 20.8 & 1.72 & 17.9 & 1.57 & 464 $\left\{
	6.1^{\circ }\right\} $ & 252 $\left\{ 6.1^{\circ }\right\} $ & 5.44 $\left\{
	33^{\circ }\right\} $ & 1.21 $\left\{ 62^{\circ }\right\} $ \\
	& 40 & 380 & 105 & 11.9 & 0.742 & 10.1 & 0.676 & 425 $\left\{ 4.6^{\circ
	}\right\} $ & 189 $\left\{ 4.7^{\circ }\right\} $ & 2.42 $\left\{ 27^{\circ
	}\right\} $ & 0.481 $\left\{ 55^{\circ }\right\} $\\
	\hline
\end{tabular}
	
}
	
\end{table*}

\begin{table*}[htbp]
	\caption{Shear modulus $G_{12}$ values for the chiral uniform tilings having different slenderness ratio $\lambda $. In the table, the values of $G_{12}$
		are multiplied by $10^{3}$ for representation purposes.}
	\label{T:4}
\centering
\resizebox{\textwidth}{!}{	

\begin{tabular}{llllllllllll}
	\hline
	& $\lambda $ & $\left[ \bar{G}_{12}\right] _{0}$ & $\left[ \tilde{G}_{12}%
	\right] _{0}$ & $\left[ \bar{G}_{12}\right] _\mathrm{UL}$ & $\left[ \tilde{G}_{12}%
	\right] _\mathrm{UL}$ & $\left[ \bar{G}_{12}\right] _\mathrm{SPL}$ & $\left[ \tilde{G}_{12}%
	\right] _\mathrm{SPL}$ & $\left[ \bar{G}_{12}\right] _{\max }\left\{ \alpha
	\right\} $ & $\left[ \tilde{G}_{12}\right] _{\max }\left\{ \alpha \right\} $
	& $\left[ \bar{G}_{12}\right] _{\min }\left\{ \alpha \right\} $ & $\left[
	\tilde{G}_{12}\right] _{\min }\left\{ \alpha \right\} $ \\
	\hline
	& 20 & 167 & 85.3 & 144 & 38.9 & 119 & 42.2 & 185 $\left\{ 9.6^{\circ
	}\right\} $ & 116 $\left\{ 9.7^{\circ }\right\} $ & 99.8 $\left\{ 28^{\circ
	}\right\} $ & 38.9 $\left\{ 120^{\circ }\right\} $ \\
	$M_{1}$ & 30 & 148 & 53.4 & 137 & 25.5 & 109 & 26.8 & 157 $\left\{
	6.4^{\circ }\right\} $ & 73.5 $\left\{ 6.5^{\circ }\right\} $ & 77.4 $%
	\left\{ 19^{\circ }\right\} $ & 25.5 $\left\{ 120^{\circ }\right\} $ \\
	& 40 & 141 & 39.1 & 134 & 19.0 & 104 & 19.7 & 146 $\left\{ 4.8^{\circ
	}\right\} $ & 54.2 $\left\{ 4.9^{\circ }\right\} $ & 66.9 $\left\{ 15^{\circ
	}\right\} $ & 19.0 $\left\{ 120^{\circ }\right\} $ \\
	&  &  &  &  &  &  &  &  &  &  &  \\
	& 20 & 16.1 & 3.06 & 16.1 & 3.06 & 29.9 & 7.18 & 52.4 $\left\{ 16^{\circ
	}\right\} $ & 18.5 $\left\{ 16^{\circ }\right\} $ & 16.1 $\left\{ 0^{\circ
	},120^{\circ }\right\} $ & 3.06 $\left\{ 0^{\circ },120^{\circ }\right\} $
	\\
	$M_{2}$ & 30 & 7.07 & 0.911 & 7.07 & 0.911 & 13.8 & 2.26 & 24.8 $\left\{
	6.6^{\circ }\right\} $ & 5.80 $\left\{ 6.7^{\circ }\right\} $ & 7.07 $%
	\left\{ 0^{\circ },120^{\circ }\right\} $ & 0.911 $\left\{ 0^{\circ
	},120^{\circ }\right\} $ \\
	& 40 & 3.94 & 0.384 & 3.94 & 0.384 & 7.78 & 0.967 & 14.3 $\left\{ 4.9^{\circ
	}\right\} $ & 2.59 $\left\{ 5.0^{\circ }\right\} $ & 3.94 $\left\{ 0^{\circ
	},120^{\circ }\right\} $ & 0.384 $\left\{ 0^{\circ },120^{\circ }\right\} $
	\\
	&  &  &  &  &  &  &  &  &  &  &  \\
	& 20 & 167 & 85.3 & 11.6 & 1.40 & 11.1 & 1.44 & 205 $\left\{ 8.9^{\circ
	}\right\} $ & 143 $\left\{ 9.0^{\circ }\right\} $ & 8.13 $\left\{ 47^{\circ
	}\right\} $ & 1.40 $\left\{ 150^{\circ }\right\} $ \\
	$M_{3}$ & 30 & 148 & 53.4 & 5.32 & 0.438 & 5.05 & 0.445 & 172 $\left\{
	6.1^{\circ }\right\} $ & 93.7 $\left\{ 6.1^{\circ }\right\} $ & 2.78 $%
	\left\{ 34^{\circ }\right\} $ & 0.438 $\left\{ 148^{\circ }\right\} $ \\
	& 40 & 141 & 39.1 & 3.01 & 0.188 & 2.84 & 0.190 & 158 $\left\{ 4.6^{\circ
	}\right\} $ & 70.2 $\left\{ 4.7^{\circ }\right\} $ & 1.28 $\left\{ 27^{\circ
	}\right\} $ & 0.188 $\left\{ 148^{\circ }\right\} $\\
	\hline
\end{tabular}

}
	
\end{table*}

\begin{table*}[htbp]
	\caption{Planar bulk modulus $K$ values for the chiral uniform tilings having different
		slenderness ratio $\lambda $. In the table, the values of $K$ are multiplied
		by $10^{3}$  for representation purposes.}
	\label{T:5}
\centering
\resizebox{\textwidth}{!}{	

\begin{tabular}{llllllllllll}
	\hline
	& $\lambda $ & $\left[ \bar{K}\right] _{0}$ & $\left[ \tilde{K}\right] _{0}$
	& $\left[ \bar{K}\right] _\mathrm{UL}$ & $\left[ \tilde{K}\right] _\mathrm{UL}$ & $\left[
	\bar{K}\right] _\mathrm{SPL}$ & $\left[ \tilde{K}\right] _\mathrm{SPL}$ & $\left[ \bar{K}%
	\right] _{\max }\left\{ \alpha \right\} $ & $\left[ \tilde{K}\right] _{\max
	}\left\{ \alpha \right\} $ & $\left[ \bar{K}\right] _{\min }\left\{ \alpha
	\right\} $ & $\left[ \tilde{K}\right] _{\min }\left\{ \alpha \right\} $ \\
	\hline
	& 20 & 345 & 176 & 291 & 78.5 & 34.9 & 12.4 & 373 $\left\{ 9.6^{\circ
	}\right\} $ & 233 $\left\{ 9.6^{\circ }\right\} $ & 23.1 $\left\{ 33^{\circ
	}\right\} $ & 10.7 $\left\{ 42^{\circ }\right\} $ \\
	$M_{1}$ & 30 & 308 & 111 & 276 & 51.5 & 13.8 & 3.41 & 324 $\left\{
	6.4^{\circ }\right\} $ & 151 $\left\{ 6.5^{\circ }\right\} $ & 6.87 $\left\{
	27^{\circ }\right\} $ & 2.48 $\left\{ 34^{\circ }\right\} $ \\
	& 40 & 292 & 80.9 & 269 & 38.4 & 7.25 & 1.37 & 301 $\left\{ 4.8^{\circ
	}\right\} $ & 112 $\left\{ 4.9^{\circ }\right\} $ & 3.04 $\left\{ 23^{\circ
	}\right\} $ & 0.898 $\left\{ 30^{\circ }\right\} $ \\
	&  &  &  &  &  &  &  &  &  &  &  \\
	& 20 & 282 & 53.6 & 282 & 53.6 & 35.8 & 8.59 & 311 $\left\{ 9.6^{\circ
	}\right\} $ & 102 $\left\{ 9.8^{\circ }\right\} $ & 28.8 $\left\{ 42^{\circ
	}\right\} $ & 7.89 $\left\{ 48^{\circ }\right\} $ \\
	$M_{2}$ & 30 & 271 & 34.9 & 271 & 34.9 & 15.0 & 2.45 & 290 $\left\{
	6.4^{\circ }\right\} $ & 67.4 $\left\{ 6.6^{\circ }\right\} $ & 10.0 $%
	\left\{ 35^{\circ }\right\} $ & 2.02 $\left\{ 42^{\circ }\right\} $ \\
	& 40 & 265 & 25.9 & 265 & 25.9 & 8.06 & 1.00 & 280 $\left\{ 4.8^{\circ
	}\right\} $ & 50.4 $\left\{ 4.9^{\circ }\right\} $ & 4.74 $\left\{ 31^{\circ
	}\right\} $ & 0.765 $\left\{ 38^{\circ }\right\} $ \\
	&  &  &  &  &  &  &  &  &  &  &  \\
	& 20 & 345 & 176 & 253 & 30.5 & 74.9 & 9.68 & 410 $\left\{ 8.9^{\circ
	}\right\} $ & 285 $\left\{ 8.9^{\circ }\right\} $ & 8.98 $\left\{ 41^{\circ
	}\right\} $ & 2.59 $\left\{ 58^{\circ }\right\} $ \\
	$M_{3}$ & 30 & 308 & 111 & 245 & 20.2 & 38.5 & 3.39 & 353 $\left\{
	6.1^{\circ }\right\} $ & 192 $\left\{ 6.2^{\circ }\right\} $ & 2.65 $\left\{
	32^{\circ }\right\} $ & 0.651 $\left\{ 50^{\circ }\right\} $ \\
	& 40 & 292 & 80.9 & 242 & 15.1 & 22.8 & 1.53 & 326 $\left\{ 4.6^{\circ
	}\right\} $ & 145 $\left\{ 4.7^{\circ }\right\} $ & 1.14 $\left\{ 27^{\circ
	}\right\} $ & 0.246 $\left\{ 45^{\circ }\right\} $\\
	\hline
\end{tabular}

}
	
\end{table*}

\begin{table*}[htbp]
	\caption{Poisson's ratio $\nu _{12}$ values for M$_4$-type uniform tilings having different slenderness ratios $\lambda$.}
	\label{T:6}
\centering
\resizebox{\textwidth}{!}{	

\begin{tabular}{llllllll}
	\hline
	& $\lambda $ &  & $\left[ \nu _{12}\right] _{0}$ $\left\langle \theta
	\right\rangle $ & $\left[ \nu _{12}\right] _\mathrm{UL}$ $\left\langle \theta
	\right\rangle $ & $\left[ \nu _{12}\right] _\mathrm{SPL}$ $\left\langle \theta
	\right\rangle $ & $\left[ \nu _{12}\right] _{\max }\left\{ \alpha \right\} $
	$\left\langle \theta \right\rangle $ & $\left[ \nu _{12}\right] _{\min
	}\left\{ \alpha \right\} $ $\left\langle \theta \right\rangle $ \\
	\hline
	& 20 & max & 0.938 $\left\langle 45.0^{\circ }\right\rangle $ & 0.955 $%
	\left\langle 45.0^{\circ }\right\rangle $ & 0.799 $\left\langle 63.0^{\circ
	}\right\rangle $ & \textbf{0.955} $\left\{ 135^{\circ }\right\} $ $%
	\left\langle 45.0^{\circ }\right\rangle $ & 0.573 $\left\{ 44^{\circ
	}\right\} $ $\left\langle 72.7^{\circ }\right\rangle $ \\
	&  & min & 0.0751 $\left\langle 0.0^{\circ }\right\rangle $ & 0.881 $%
	\left\langle 0.0^{\circ }\right\rangle $ & 0.146 $\left\langle 18.0^{\circ
	}\right\rangle $ & 0.881 $\left\{ 135^{\circ }\right\} $ $\left\langle
	0.0^{\circ }\right\rangle $ & \textbf{-0.820} $\left\{ 44^{\circ }\right\} $
	$\left\langle 27.7^{\circ }\right\rangle $ \\
	$\nu _{12}$ & 30 & max & 0.973 $\left\langle 45.0^{\circ }\right\rangle $ &
	0.979 $\left\langle 45.0^{\circ }\right\rangle $ & 0.814 $\left\langle
	63.3^{\circ }\right\rangle $ & \textbf{0.979} $\left\{ 135^{\circ }\right\} $
	$\left\langle 45.0^{\circ }\right\rangle $ & 0.570 $\left\{ 44^{\circ
	}\right\} $ $\left\langle 72.6^{\circ }\right\rangle $ \\
	&  & min & 0.0496 $\left\langle 0.0^{\circ }\right\rangle $ & 0.946 $%
	\left\langle 0.0^{\circ }\right\rangle $ & 0.147 $\left\langle 18.3^{\circ
	}\right\rangle $ & 0.946 $\left\{ 135^{\circ }\right\} $ $\left\langle
	0.0^{\circ }\right\rangle $ & \textbf{-0.929} $\left\{ 44^{\circ }\right\} $
	$\left\langle 27.6^{\circ }\right\rangle $ \\
	& 40 & max & 0.985 $\left\langle 45.0^{\circ }\right\rangle $ & 0.988 $%
	\left\langle 45.0^{\circ }\right\rangle $ & 0.818 $\left\langle 63.5^{\circ
	}\right\rangle $ & \textbf{0.988} $\left\{ 135^{\circ }\right\} $ $%
	\left\langle 45.0^{\circ }\right\rangle $ & 0.566 $\left\{ 44^{\circ
	}\right\} $ $\left\langle 72.6^{\circ }\right\rangle $ \\
	&  & min & 0.0371 $\left\langle 0.0^{\circ }\right\rangle $ & 0.969 $%
	\left\langle 0.0^{\circ }\right\rangle $ & 0.148 $\left\langle 18.5^{\circ
	}\right\rangle $ & 0.969 $\left\{ 135^{\circ }\right\} $ $\left\langle
	0.0^{\circ }\right\rangle $ & \textbf{-0.962} $\left\{ 44^{\circ }\right\} $
	$\left\langle 27\text{.}6^{\circ }\right\rangle $\\
	\hline
\end{tabular}
	
}
	
\end{table*}

\begin{table*}[htbp]
	\caption{Young's modulus $E_{1}$ values for M$_4$-type uniform tilings
		having different slenderness ratios $\lambda $. In the table, the values of $%
		E_{1}$ are multiplied by $10^{3}$  for representation purposes.}
\label{T:7}

\centering
\resizebox{\textwidth}{!}{	

\begin{tabular}{llllllll}
	\hline
	& $\lambda $ &  & $\left[ E_{1}\right] _{0}$ & $\left[ E_{1}\right] _\mathrm{UL}$ &
	$\left[ E_{1}\right] _\mathrm{SPL}$ & $\left[ E_{1}\right] _{\max }\left\{ \alpha
	\right\} $ $\left\langle \theta \right\rangle $ & $\left[ E_{1}\right]
	_{\min }\left\{ \alpha \right\} $ $\left\langle \theta \right\rangle $ \\
	\hline
	& 20 & max & 562 $\left\langle 0.0^{\circ }\right\rangle $ & 64.3 $%
	\left\langle 0.0^{\circ }\right\rangle $ & 73.0 $\left\langle 18.0^{\circ
	}\right\rangle $ & \textbf{609} $\left\{ 9.2^{\circ }\right\} $ $%
	\left\langle 0.3^{\circ }\right\rangle $ & 55.6 $\left\{ 49^{\circ }\right\}
	$ $\left\langle 26.6^{\circ }\right\rangle $ \\
	&  & min & 37.9 $\left\langle 45.0^{\circ }\right\rangle $ & 24.3 $%
	\left\langle 45.0^{\circ }\right\rangle $ & 17.2 $\left\langle 63.0^{\circ
	}\right\rangle $ & 119 $\left\{ 9.2^{\circ }\right\} $ $\left\langle
	45.3^{\circ }\right\rangle $ & \textbf{12.5} $\left\{ 49^{\circ }\right\} $ $%
	\left\langle 71.6^{\circ }\right\rangle $ \\
	$\bar{E}_{1}$ & 30 & max & 540 $\left\langle 0.0^{\circ }\right\rangle $ &
	28.3 $\left\langle 0.0^{\circ }\right\rangle $ & 33.3 $\left\langle
	18.3^{\circ }\right\rangle $ & \textbf{573} $\left\{ 6.2^{\circ }\right\} $ $%
	\left\langle 0.1^{\circ }\right\rangle $ & 18.9 $\left\{ 40^{\circ }\right\}
	$ $\left\langle 28.5^{\circ }\right\rangle $ \\
	&  & min & 15.6 $\left\langle 45.0^{\circ }\right\rangle $ & 10.8 $%
	\left\langle 45.0^{\circ }\right\rangle $ & 7.28 $\left\langle 63.3^{\circ
	}\right\rangle $ & 53.0 $\left\{ 6.2^{\circ }\right\} $ $\left\langle
	45.1^{\circ }\right\rangle $ & \textbf{4.43} $\left\{ 40^{\circ }\right\} $ $%
	\left\langle 73.5^{\circ }\right\rangle $ \\
	& 40 & max & 529 $\left\langle 0.0^{\circ }\right\rangle $ & 15.7 $%
	\left\langle 0.0^{\circ }\right\rangle $ & 18.8 $\left\langle 18.5^{\circ
	}\right\rangle $ & \textbf{555} $\left\{ 4.7^{\circ }\right\} $ $%
	\left\langle 0.1^{\circ }\right\rangle $ & 8.47 $\left\{ 34^{\circ }\right\}
	$ $\left\langle 29.7^{\circ }\right\rangle $ \\
	&  & min & 8.44 $\left\langle 45.0^{\circ }\right\rangle $ & 6.02 $%
	\left\langle 45.0^{\circ }\right\rangle $ & 3.99 $\left\langle 63.5^{\circ
	}\right\rangle $ & 29.8 $\left\{ 4.7^{\circ }\right\} $ $\left\langle
	45.1^{\circ }\right\rangle $ & \textbf{2.14} $\left\{ 34^{\circ }\right\} $ $%
	\left\langle 74.7^{\circ }\right\rangle $ \\
	&  &  &  &  &  &  &  \\
	& 20 & max & 178 $\left\langle 0.0^{\circ }\right\rangle $ & 10.9 $%
	\left\langle 0.0^{\circ }\right\rangle $ & 14.3 $\left\langle 18.0^{\circ
	}\right\rangle $ & \textbf{300} $\left\{ 9.3^{\circ }\right\} $ $%
	\left\langle 0.3^{\circ }\right\rangle $ & 15.4 $\left\{ 73^{\circ }\right\}
	$ $\left\langle 21.8^{\circ }\right\rangle $ \\
	&  & min & 12.0 $\left\langle 45.0^{\circ }\right\rangle $ & 4.10 $%
	\left\langle 45.0^{\circ }\right\rangle $ & 3.37 $\left\langle 63.0^{\circ
	}\right\rangle $ & 59.1 $\left\{ 9.3^{\circ }\right\} $ $\left\langle
	45.3^{\circ }\right\rangle $ & \textbf{3.23} $\left\{ 73^{\circ }\right\} $ $%
	\left\langle 66.8^{\circ }\right\rangle $ \\
	$\tilde{E}_{1}$ & 30 & max & 118 $\left\langle 0.0^{\circ }\right\rangle $ &
	3.24 $\left\langle 0.0^{\circ }\right\rangle $ & 4.44 $\left\langle
	18.3^{\circ }\right\rangle $ & \textbf{209} $\left\{ 6.3^{\circ }\right\} $ $%
	\left\langle 0.1^{\circ }\right\rangle $ & 4.62 $\left\{ 65^{\circ }\right\}
	$ $\left\langle 23.6^{\circ }\right\rangle $ \\
	&  & min & 3.40 $\left\langle 45.0^{\circ }\right\rangle $ & 1.23 $%
	\left\langle 45.0^{\circ }\right\rangle $ & 0.971 $\left\langle 63.3^{\circ
	}\right\rangle $ & 19.6 $\left\{ 6.3^{\circ }\right\} $ $\left\langle
	45.1^{\circ }\right\rangle $ & \textbf{0.891} $\left\{ 65^{\circ }\right\} $
	$\left\langle 68.6^{\circ }\right\rangle $ \\
	& 40 & max & 87.7 $\left\langle 0.0^{\circ }\right\rangle $ & 1.36 $%
	\left\langle 0.0^{\circ }\right\rangle $ & 18.9 $\left\langle 18.5^{\circ
	}\right\rangle $ & \textbf{160} $\left\{ 4.7^{\circ }\right\} $ $%
	\left\langle 0.1^{\circ }\right\rangle $ & 1.86 $\left\{ 34^{\circ }\right\}
	$ $\left\langle 24.7^{\circ }\right\rangle $ \\
	&  & min & 1.40 $\left\langle 45.0^{\circ }\right\rangle $ & 0.522 $%
	\left\langle 45.0^{\circ }\right\rangle $ & 0.403 $\left\langle 63.5^{\circ
	}\right\rangle $ & 8.79 $\left\{ 4.7^{\circ }\right\} $ $\left\langle
	45.1^{\circ }\right\rangle $ & \textbf{0.359} $\left\{ 34^{\circ }\right\} $
	$\left\langle 69.7^{\circ }\right\rangle $\\
	\hline
\end{tabular}

}
	
\end{table*}
	
\begin{table*}[htbp]
	\caption{Shear Modulus $G_{12}$ values for M$_4$-type uniform tilings having
		different slenderness ratios $\lambda $. In the table, the values of $G_{12}$
		are multiplied by $10^{3}$  for representation purposes.}
\label{T:8}
\centering
\resizebox{\textwidth}{!}{	

\begin{tabular}{llllllll}
	\hline
	& $\lambda $ &  & $\left[ G_{12}\right] _{0}$ & $\left[ G_{12}\right] _\mathrm{UL}$
	& $\left[ G_{12}\right] _\mathrm{SPL}$ & $\left[ G_{12}\right] _{\max }\left\{
	\alpha \right\} $ $\left\langle \theta \right\rangle $ & $\left[ G_{12}%
	\right] _{\min }\left\{ \alpha \right\} $ $\left\langle \theta \right\rangle
	$ \\
	\hline
	& 20 & max & 261 $\left\langle 45.0^{\circ }\right\rangle $ & 17.1 $%
	\left\langle 45.0^{\circ }\right\rangle $ & 31.9 $\left\langle 63.0^{\circ
	}\right\rangle $ & \textbf{270} $\left\{ 9.2^{\circ }\right\} $ $%
	\left\langle 45.3^{\circ }\right\rangle $ & 127 $\left\{ 53^{\circ }\right\}
	$ $\left\langle 70.8^{\circ }\right\rangle $ \\
	&  & min & 9.78 $\left\langle 0.0^{\circ }\right\rangle $ & 6.21 $%
	\left\langle 0.0^{\circ }\right\rangle $ & 4.78 $\left\langle 18.0^{\circ
	}\right\rangle $ & 32.5 $\left\{ 9.2^{\circ }\right\} $ $\left\langle
	0.3^{\circ }\right\rangle $ & \textbf{3.89} $\left\{ 53^{\circ }\right\} $ $%
	\left\langle 25.8^{\circ }\right\rangle $ \\
	$\bar{G}_{12}$ & 30 & max & 257 $\left\langle 45.0^{\circ }\right\rangle $ &
	7.27 $\left\langle 45.0^{\circ }\right\rangle $ & 14.5 $\left\langle
	63.3^{\circ }\right\rangle $ & \textbf{263} $\left\{ 6.2^{\circ }\right\} $ $%
	\left\langle 45.1^{\circ }\right\rangle $ & 137 $\left\{ 43^{\circ }\right\}
	$ $\left\langle 72.8^{\circ }\right\rangle $ \\
	&  & min & 3.95 $\left\langle 0.0^{\circ }\right\rangle $ & 2.72 $%
	\left\langle 0.0^{\circ }\right\rangle $ & 2.01 $\left\langle 18.3^{\circ
	}\right\rangle $ & 13.8 $\left\{ 6.2^{\circ }\right\} $ $\left\langle
	0.1^{\circ }\right\rangle $ & \textbf{1.42} $\left\{ 43^{\circ }\right\} $ $%
	\left\langle 27.8^{\circ }\right\rangle $ \\
	& 40 & max & 255 $\left\langle 45.0^{\circ }\right\rangle $ & 3.99 $%
	\left\langle 45.0^{\circ }\right\rangle $ & 8.17 $\left\langle 63.5^{\circ
	}\right\rangle $ & \textbf{260} $\left\{ 4.7^{\circ }\right\} $ $%
	\left\langle 45.1^{\circ }\right\rangle $ & 60.8 $\left\{ 36^{\circ
	}\right\} $ $\left\langle 74.3^{\circ }\right\rangle $ \\
	&  & min & 2.13 $\left\langle 0.0^{\circ }\right\rangle $ & 1.51 $%
	\left\langle 0.0^{\circ }\right\rangle $ & 1.10 $\left\langle 18.5^{\circ
	}\right\rangle $ & 7.65 $\left\{ 4.7^{\circ }\right\} $ $\left\langle
	0.1^{\circ }\right\rangle $ & \textbf{0.703} $\left\{ 36^{\circ }\right\} $ $%
	\left\langle 29.3^{\circ }\right\rangle $ \\
	&  &  &  &  &  &  &  \\
	& 20 & max & 82.7 $\left\langle 45.0^{\circ }\right\rangle $ & 2.89 $%
	\left\langle 45.0^{\circ }\right\rangle $ & 6.24 $\left\langle 63.0^{\circ
	}\right\rangle $ & \textbf{133} $\left\{ 9.3^{\circ }\right\} $ $%
	\left\langle 45.3^{\circ }\right\rangle $ & 8.02 $\left\{ 83^{\circ
	}\right\} $ $\left\langle 64.7^{\circ }\right\rangle $ \\
	&  & min & 3.09 $\left\langle 0.0^{\circ }\right\rangle $ & 1.05 $%
	\left\langle 0.0^{\circ }\right\rangle $ & 0.936 $\left\langle 18.0^{\circ
	}\right\rangle $ & 16.2 $\left\{ 9.3^{\circ }\right\} $ $\left\langle
	0.3^{\circ }\right\rangle $ & \textbf{0.931} $\left\{ 83^{\circ }\right\} $ $%
	\left\langle 19.7^{\circ }\right\rangle $ \\
	$\tilde{G}_{12}$ & 30 & max & 56.0 $\left\langle 45.0^{\circ }\right\rangle $
	& 0.834 $\left\langle 45.0^{\circ }\right\rangle $ & 1.93 $\left\langle
	63.3^{\circ }\right\rangle $ & \textbf{95.9} $\left\{ 6.3^{\circ }\right\} $
	$\left\langle 45.1^{\circ }\right\rangle $ & 3.61 $\left\{ 76^{\circ
	}\right\} $ $\left\langle 66.4^{\circ }\right\rangle $ \\
	&  & min & 0.861 $\left\langle 0.0^{\circ }\right\rangle $ & 0.312 $%
	\left\langle 0.0^{\circ }\right\rangle $ & 0.268 $\left\langle 18.3^{\circ
	}\right\rangle $ & 5.12 $\left\{ 6.3^{\circ }\right\} $ $\left\langle
	0.1^{\circ }\right\rangle $ & \textbf{0.262} $\left\{ 76^{\circ }\right\} $ $%
	\left\langle 21.4^{\circ }\right\rangle $ \\
	& 40 & max & 42.4 $\left\langle 45.0^{\circ }\right\rangle $ & 0.346 $%
	\left\langle 45.0^{\circ }\right\rangle $ & 0.825 $\left\langle 63.5^{\circ
	}\right\rangle $ & \textbf{75.0} $\left\{ 4.8^{\circ }\right\} $ $%
	\left\langle 45.1^{\circ }\right\rangle $ & 2.02 $\left\{ 72^{\circ
	}\right\} $ $\left\langle 67.3^{\circ }\right\rangle $ \\
	&  & min & 0.352 $\left\langle 0.0^{\circ }\right\rangle $ & 0.131 $%
	\left\langle 0.0^{\circ }\right\rangle $ & 0.111 $\left\langle 18.5^{\circ
	}\right\rangle $ & 2.25 $\left\{ 4.8^{\circ }\right\} $ $\left\langle
	0.1^{\circ }\right\rangle $ & \textbf{0.107} $\left\{ 72^{\circ }\right\} $ $%
	\left\langle 22.3^{\circ }\right\rangle $\\
	\hline
\end{tabular}
	
}

\end{table*}

\begin{table*}[htbp]
	\caption{Planar bulk Modulus $K$ values for M$_4$-type uniform tilings having different slenderness ratios $\lambda $. In the table, the values of $K$ are multiplied by $10^{3}$  for representation purposes.}
\label{T:9}

\centering	
\begin{tabular}{lllllll}
	\hline
	& $\lambda $ & $\left[ K\right] _{0}$ & $\left[ K\right] _\mathrm{UL}$ & $\left[ K
	\right] _\mathrm{SPL}$ & $\left[ K\right] _{\max }\left\{ \alpha \right\} $ & $
	\left[ K\right] _{\min }\left\{ \alpha \right\} $ \\
	\hline
	& 20 & 304 & 271 & 42.7 & 349 $\left\{ 9.3^{\circ }\right\} $ & 14.8 $%
	\left\{ 41^{\circ }\right\} $ \\
	$\bar{K}$ & 30 & 284 & 261 & 19.5 & 314 $\left\{ 6.3^{\circ }\right\} $ &
	4.75 $\left\{ 34^{\circ }\right\} $ \\
	& 40 & 275 & 256 & 11.0 & 297 $\left\{ 4.8^{\circ }\right\} $ & 2.16 $%
	\left\{ 29^{\circ }\right\} $ \\
	& 20 & 96.1 & 45.7 & 8.37 & 172 $\left\{ 9.4^{\circ }\right\} $ & 4.58 $%
	\left\{ 52^{\circ }\right\} $ \\
	$\tilde{K}$ & 30 & 61.8 & 29.9 & 2.60 & 115 $\left\{ 6.4^{\circ }\right\} $
	& 1.15 $\left\{ 44^{\circ }\right\} $ \\
	& 40 & 45.6 & 22.2 & 1.11 & 86.1 $\left\{ 4.8^{\circ }\right\} $ & 0.433 $%
	\left\{ 40^{\circ }\right\} $\\
	\hline
\end{tabular}
\end{table*}

\clearpage
\bibliography{bibliography}

\end{document}